\documentclass[11pt]{JHEP3}
\usepackage{epsfig,multicol,bbm}
\usepackage{cite}
\usepackage{amsmath,amsfonts,amssymb,longtable}
\usepackage{array} 
\usepackage{graphicx}
\usepackage{subfig}
\input epsf.sty

\newcommand{\bright}{\begin{flushright}}
\newcommand{\eright}{\end{flushright}}
\newcommand{\bminip}{\begin{minipage}}
\newcommand{\eminip}{\end{minipage}}
\newcommand{\bcent}{\begin{center}}
\newcommand{\ecent}{\end{center}}

\newcommand{\E}[1]{\ensuremath{\mathrm{E}_{#1}}} 

\newcommand{\SO}[1]{\ensuremath{\mathrm{SO}(#1)}}
\newcommand{\SU}[1]{\ensuremath{\mathrm{SU}(#1)}}
\newcommand{\U}[1]{\ensuremath{\mathrm{U}(#1)}}
\newcommand{\x}{\ensuremath{\times}}
\newcommand{\Z}[1]{\ensuremath{\mathbbm{Z}_{#1}}} 
\newcommand{\I}{\mathrm{i}}
\newcommand{\id}{\mathbbm{1}}
\newcommand{\bs}[1]{\ensuremath{\boldsymbol{#1}}}
\newcommand{\bsb}[1]{\ensuremath{\boldsymbol{\bar{#1}}}}
\newcommand{\ket}[1]{|#1\rangle}

\def\be{\begin{equation}}
\def\ee{\end{equation}}
\def\bea{\begin{eqnarray}}
\def\eea{\end{eqnarray}}

\def\nn{\nonumber}

\def\p{\phi}

\def\I{\mathrm{i}}

\title{\Huge On Moduli Stabilisation and de Sitter Vacua in  MSSM Heterotic Orbifolds}

\author{Susha L.~Parameswaran$^a$, Sa\'ul Ramos-S\'anchez$^{b}$, Ivonne Zavala$^c$ \\ \\
${}^a$Department of Physics and Astronomy, Uppsala University, 
  P.O. Box 803, S-75108, Uppsala, Sweden\\ 
${}^b$Deutsches Elektronen-Synchrotron DESY, Hamburg, Germany\\
${}^c$Bethe Center for Theoretical Physics and
  Physikalisches Institut der Universit\"at Bonn,
  Nussallee 12, 53115 Bonn, Germany\\ \\
E-mail: \email{susha.parameswaran@fysast.uu.se}\,, 
\email{ramos@mail.desy.de}\,, 
\email{zavala@th.physik.uni-bonn.de}
}

\preprint{DESY 10-148}

\abstract{ We study the problem of moduli stabilisation in explicit heterotic
orbifold compactifications, whose spectra contain the MSSM plus some
vector-like exotics that can be decoupled.  Considering {\it all} the
bulk moduli, we obtain the 4D low energy effective action for the
compactification, which has contributions from various, computable,
perturbative and non-perturbative effects. Hidden sector gaugino
condensation and string worldsheet instantons result in a combination of
racetrack, KKLT-like and cusp-form contributions to the superpotential, which
lift all the bulk moduli directions.  We point out the properties
observed in our concrete models, which tend to be missed when only
``generic'' features of a model are assumed. We search for interesting
vacua and find several de Sitter solutions, but {\it so far}, they all turn
out to be unstable.}


\begin{document}

\section{Introduction}

In searching for the Standard Model (SM) of particle physics and its
supersymmetric generalisations within String Theory, a myriad of
constructions have been explored.  Promising models have been built
using heterotic orbifold compactifications, intersecting D-branes or
D-branes at singularities and F-theory GUTs, to name a few.  What all
these models have in common, is that progress was made by focussing on
the realisation of phenomenologically viable gauge groups and chiral
matter spectra and, in particular, by neglecting gravity and the
dynamics of the compactification.

Eventually, however, it will be necessary to embed the supersymmetric
standard model (SSM) into globally consistent and stable string theory
constructions.  For example, supersymmetry breaking is usually
associated with the dynamics of the moduli, and therefore in order to
make predictions for the soft masses we require an understanding of
how the moduli are stabilised.
Ideas abound also in how to address the
moduli stabilisation problem, and some progress has also been made towards 
dove-tailing these constructions with those of the SSM, particularly in Type 
IIB models \cite{realmodstabII}.  So far though, the possibility of realising
the SSM in a genuine metastable string compactification seems out of reach.

In this paper we continue towards this objective within the context of
heterotic orbifold compactifications.  Heterotic orbifolds are
arguably amongst the simplest frameworks for such studies.  They have
had a long history as candidates for realistic string theory
constructions, particularly in the eighties and nineties, and as such,
much effort was also put towards understanding supersymmetry breaking
and moduli stabilisation there.  Recent years have seen a significant
revival in this class of models, since it was realised that the idea
of orbifold GUTs applied to heterotic orbifolds makes an effective
tool to uncover the MSSM in the latter \cite{BHLR}.

This progress is marked by the studies of a particularly fertile patch
in the heterotic landscape, in which around 300
models were found
containing the MSSM spectrum, together with only some vector-like
exotics that can be decoupled \cite{Saul}.  They are built using a
\Z6--II orbifold compactification, with a non-standard gauge embedding
(leading to ${\mathcal{N}}=(0,2)$ worldsheet supersymmetry) and some
number of discrete Wilson lines.  We shall consider the problem of
moduli stabilisation in some of these concrete promising models.

In their totality, these models contain, as well as the MSSM and the
supergravity multiplets, a hidden gauge sector, a number of vector-like
exotics, SM singlets (typically charged at least under Abelian factors 
of the hidden gauge sector), and the bulk moduli.  Amongst the twisted
fields there are also flat directions, and some of these may have a
geometric interpretation as blow-up modes.  The latter are known as
twisted moduli, but, interestingly for the possibility that our
universe is an orbifold, they are not always present \cite{DFMS}. 
As a first step, our focus will be on all 
the bulk moduli.  Therefore, although we will begin by treating all
degrees of freedom, at the end we shall assume that the vector-like
exotics and SM singlets are decoupled at some scale
close to the string scale, consistently with supersymmetry, {\it e.g.}
due to dynamics such as those discussed in \cite{Saul}.

In order to study the bulk moduli dynamics, it is first necessary to
construct the low energy effective field theory for the given orbifold
compactification.  Fortunately, there is a good knowledge of the
effective Lagrangian for orbifold compactifications (for some nice
reviews see \cite{BLreview,Choi:2006qh}).  The relevant quantities can
be computed using dimensional reduction and conformal field theory
techniques \cite{yukawas,DKL90}.  Another powerful tool \cite{FLST,Nillesduality,IL}
is modular invariance, which in the simplest models corresponds to an
$SL(2, \Z{})$ symmetry for each geometric modulus, and which is
expected to hold even non-perturbatively. 
In what follows we shall
compute the leading terms in the low energy effective field theory
that describes concrete MSSM orbifold compactifications.
Then, building on the work 
of the eighties and nineties, since all directions can be lifted, it seems natural to expect that (meta)stable vacua can exist.

Indeed, although the dilaton and geometric moduli are all flat
directions at tree level, non-perturbative effects tend to lift them
all.  Since the dilaton describes the tree level gauge couplings,
gaugino condensation in a hidden non-Abelian sector leads to a
non-trivial potential for that field~\cite{GC}.  A racetrack 
potential~\cite{CLMR, CCM,racetrack} or KKLT-type 
potential~\cite{KKLT, saulW0, DRW} may then be sufficient to 
stabilise it.  The gauge couplings
also receive threshold corrections from massive string states that
depend, via certain cusp-forms, on several of the K\"ahler moduli and
all the complex structure moduli \cite{DKL90, Kaplunovsky:1995jw}, and
thus a gaugino condensate may also stabilise those fields or even
force compactification \cite{FILQ,CFILQ,CCM,BLST1}.  Yukawa couplings
between twisted fields arise thanks to worldsheet instantons
\cite{yukawas}, and are suppressed with the K\"ahler moduli that
describe their separation.  This lifts all the remaining K\"ahler
moduli directions \cite{LM}.  The most complete and general analysis
to date on some of these effects can be found in the seminal work
\cite{CCM}, which showed the success of racetrack dilaton
stabilisation in the presence of hidden matter, together with the
stabilisation of a universal K\"ahler modulus.  Another interesting 
explicit analysis was performed in \cite{BLST1}, where it was shown that a single gaugino
condensate is sufficient to stabilise several geometrical moduli via a
cusp-form mechanism,
although the  vacuum expectation value ({\it{vev}}) of the dilaton was  set by hand.

The present work is a  first attempt at treating all the bulk moduli together  in explicit models that give rise to the MSSM. Our models have three K\"ahler moduli, one complex structure modulus plus
the dilaton, giving a total of 10 real degrees of freedom.  The
various ingredients just described, give rise to a superpotential for
all the fields, composed of a mixture of three types of terms, which
have often been used in three popular stabilisation mechanisms:
racetrack, KKLT and cusp-form stabilisation.

Several differences occur when considering concrete models compared to
the inspiring toy models of the past, however.  Some of these are:
{\it (i)} the terms in the action are almost completely determined
with very few free parameters (which come mainly from the little studied 
physics that decouples exotics), {\it (ii)} the modular symmetry
groups are generically broken from $SL(2,\Z{})$ to some
congruence subgroup due to the presence of Wilson lines
\cite{modWL,modWLBL}, {\it (iii)} it is typically difficult to find
more than one condensing gauge group, especially without decoupling hidden matter
(since the latter tends to destroy asymptotic freedom), and
subsequently, it is hard (but not impossible) to find dilatonic
racetrack models, {\it (iv)} the moduli-dependent threshold
corrections to the gauge couplings often do not appear with the
required sign to force compactification, {\it (v)} it is not 
justified to take a universal K\"ahler modulus, and
moreover the dynamics of all the bulk moduli are highly coupled
together.  All these features make the
search for a metastable (de Sitter) vacuum more challenging than previously thought and our results underline this.

Due to the complexity of the system, our search for minima is largely
numerical.  We do find many de Sitter vacua, but all of them have
unstable directions.  We study their properties, including the
presence of almost flat directions.  Whether or not a metastable de
Sitter vacuum might be waiting to be found is currently under
investigation.

The organisation of the paper is as follows.  In the next section, we
review briefly the orbifold construction in heterotic strings. Expert
orbifolders can skip this introduction, and go directly to Section
\ref{Sec3}, where we describe the low energy effective theory arising
from generic heterotic orbifolds, which takes the form of an
${\mathcal N}=1$ supergravity.  In particular, we recount the various
perturbative and non-perturbative effects that can contribute to the
dynamics of the moduli.  Building on this discussion, Section
\ref{Sec4} is devoted to the detailed study of the problem of moduli
stabilisation in two concrete \Z6--II orbifold models, which contain
the MSSM.  We present explicit de Sitter vacua.  Finally in Section \ref{Concl}
we discuss our results and the prospects to resolve the important open
questions.

Throughout the paper we take $\kappa_4=1$ and use dimensionless
quantities. Whenever units are necessary, we say so explicitly.

\section{An Invitation to Heterotic Orbifolds}
\label{Sec2}

In this section we review briefly  the compactification of the
\E8\x\E8 heterotic string on non-freely acting, symmetric \Z{N}
toroidal orbifolds~\cite{Choi:2006qh,BLreview}, which are the
setting for our investigation.

An orbifold is obtained by modding out a discrete set of isometries
$P$ of a 6D torus $T^6$, characterised by a lattice $\Lambda$. $P$ is
called the {\it point group} and we assume the torus $T^6$ to be
factorisable, {\it i.e.} $T^6=T^2\times T^2\times T^2$. In general, the
basis vectors $e_\alpha$ of the lattice $\Lambda$ can be chosen to be
the simple roots of a Lie group, whose metric is
$g_{\alpha\beta}=e_\alpha\cdot e_\beta$. The action of the point group
generator $\vartheta$ on $\Lambda$ is given by
\begin{equation}
  \label{eq:PonLambda}
  \vartheta\,:~e_\alpha \rightarrow e_\beta \ \Theta_{\beta\alpha}\,,
\end{equation}
where $\Theta$ is the Coxeter element of the corresponding Lie
algebra. In terms of the lattice metric, Eq.~\eqref{eq:PonLambda}
becomes
\begin{equation}
  \label{eq:Pong}
  \vartheta\,:~g_{\alpha\beta}\rightarrow \Theta^T g_{\alpha\beta} \Theta\,.
\end{equation}
An admissible lattice should be invariant under $P$. This restricts
the choice of $\Lambda$, but leaves some free parameters, which
correspond to the geometric moduli of the compactification.

In \Z{N} orbifolds, the generator $\vartheta$ can be made diagonal, such that
$$\vartheta=\mathrm{diag}(e^{2\pi\I v_1},\,e^{2\pi\I v_2},\,e^{2\pi\I v_3})$$
where $v=(0,\,v_1,\,v_2,\,v_3)$ is the twist vector that parametrises
the action of the orbifold on the compact space~\footnote{Here $v_0=0$
  represents the trivial action of the twist on the Minkowski 4D space
  with complex index $\mu=0$ in light-cone gauge coordinates.}.
Since $\vartheta$ is of order $N$, it
follows that $N\,v_i = 0\mod 1$. Additionally, $v$ is subject to the
constraint $\sum_i v_i = 0$ that guarantees the holonomy to be a
(discrete) subgroup of \SU3 (but larger than \SU2) and consequently,
$\mathcal{N}=1$ in 4D.

In the bosonic formulation, points of the torus are given by six real
or, equivalently, three complex coordinates $Z_i$, $i=1,2,3$.  Under
$\vartheta$, the torus coordinates transform as
\begin{equation}
  \label{eq:PonZ}
  Z_i \stackrel{\vartheta}{\longrightarrow} \vartheta_i Z_i\,.
\end{equation}
In order to include the fact that $Z$ is a point on $T^6$, {\it i.e.} $Z \equiv Z + \lambda$ with
$\lambda\in\Lambda$, it is customary to define the {\it space group} $S$ as the semidirect product
$S=P\ltimes\Lambda$. An arbitrary space group element $g=(\vartheta^k,\,n_\alpha e_\alpha)$ acts then on
the coordinates of the compact space as
\begin{equation}
  \label{eq:SonZ}
  Z \stackrel{g}{\longrightarrow} \vartheta^k Z + n_\alpha e_\alpha,\quad\qquad n_\alpha\in\Z{}\,.
\end{equation}
It is easy to see that the action of a non-trivial space group element leaves invariant the points $f$
satisfying
\begin{equation}
  \label{eq:fp}
  (\id-\vartheta^k)\,f = n_\alpha e_\alpha,\quad\qquad n_\alpha\in\Z{}\,.
\end{equation}
Such points  are called {\it fixed points}. Sometimes, instead of $f$, one uses the associated space group
element $g$ to refer to the fixed points. Since all elements from the {\it conjugacy class} of $g$,
defined as $\{h\,g\,h^{-1}\,|\,h\in S\}$, have an equivalent action on the compact space, then the
corresponding fixed points are equivalent in the orbifold.

For each point group element $\vartheta^k$, one obtains a different number of inequivalent fixed points.
Therefore, usually one refers collectively to the set of fixed points for a given $k$ as the
$\vartheta^k$ {\it twisted sector}. For $k=0$, the twist action is trivial and therefore there are no
fixed points. This sector is called the {\it untwisted sector}.
The action of the point group on the $\vartheta^k$ twisted sector is conveniently
encoded in the {\it local twist vector} $\bs{k}=k\,v$. In orbifolds with non-prime $N$, there are some
sectors in which the twist acts trivially on one of the three two-torii. In these cases, we get {\it fixed
torii} instead of fixed points. The holonomy in these sectors is a subgroup of \SU2,
which leaves $\mathcal{N}=2$ supersymmetry unbroken.

Modular invariance requires the action of the space group to be accompanied by a corresponding action on
the 16 gauge degrees of freedom $X_I$ of the heterotic string. The action of $\vartheta$ can be described by a
shift~\footnote{Another admissible embedding is a 16D rotation in the gauge degrees of freedom.}
\begin{equation}
  \label{eq:VonX}
   X \stackrel{\vartheta}{\longrightarrow} X + \pi V\,,
\end{equation}
where $V$ is the {\it shift vector}. The lattice translations
$e_\alpha$ are translated to the gauge degrees of freedom as discrete Wilson lines 
${\cal A}_\alpha$. Thus, an arbitrary space group element
$g=(\vartheta^k,\,n_\alpha e_\alpha)$ affects the gauge degrees of freedom as
\begin{equation}
  \label{eq:gonX}
   X \stackrel{g}{\longrightarrow} X + \pi (k\, V +  n_\alpha {\cal A}_\alpha)\,.
\end{equation}
Not every shift and Wilson lines are compatible with the space group. First, from
$\vartheta^N=\id$ it follows that $N\,V$ must be a vector in the weight lattice of \E8\x\E8, {\it i.e.} the
shift has to be of order $N$ too. Also the Wilson lines are constrained.  Since the space group elements
$g=(\id,\,n_\alpha e_\alpha)$ and $\vartheta^\ell\,g\,\vartheta^{-\ell}=(\id,\,\vartheta^\ell n_\alpha
e_\alpha)\equiv(\id,\,m_\beta e_\beta)$ are in the same conjugacy class, the Wilson lines $n_\alpha
{\cal A}_\alpha$ and $m_\beta {\cal A}_\beta$ must coincide. Direct consequences of this are that not all the Wilson
lines are non-trivial nor independent, they can take only quantised values, and their order $N_\alpha$ is
restricted by the geometry of the compact space.

Demanding modular invariance of the one-loop vacuum-to-vacuum amplitudes imposes additional constraints on the
orbifold parameters which, for \Z{N} orbifolds without Wilson lines, read~\cite{ModInv} 
\begin{equation}
\label{eq:ModInvV}
N(V^2-v^2)=0\mod 2\,.
\end{equation}
In the presence of non-trivial Wilson lines, Eq.~\eqref{eq:ModInvV} has to be amended
by~\cite{Ploger:2007iq} (no summation implied)
\begin{subequations}\label{eq:ModInvA}
\begin{eqnarray}
  \label{eq:fsmiVAa}
  N_\alpha\,{\cal A}_{\alpha}\cdot V  & = & 0 \mod 2\,, \\
  \label{eq:fsmiAaAa}
  N_\alpha\,{\cal A}_{\alpha}^2  & = & 0 \mod 2\,, \\
  \label{eq:fsmiAaAb}
  \gcd(N_{\alpha},N_{\beta})\,{\cal A}_{\alpha}\cdot {\cal A}_{\beta}  & = & 0 \mod 2 \qquad (\alpha \neq \beta)\,,
\end{eqnarray}
\end{subequations}
where $\gcd$ stands for the greatest common denominator.
These conditions guarantee anomaly freedom of the low energy effective theory.

The simplest gauge embedding of the orbifold action corresponds to
identifying the shift vector $V$ with the twist vector $v$ and setting
${\cal A}_\alpha=0$ or, in the context of supergravity, to embed the spin
connection into the gauge connection.  This is the trivial solution to
the modular invariance condition~\eqref{eq:ModInvV} and it is called
{\it standard embedding} (SE).  On the string theory side, the
standard embedding preserves $(2,2)$ worldsheet supersymmetry.  From
the phenomenological point of view, the SE is not very attractive,
since it is not possible to get realistic spectra~\footnote{There are
  two reasons for this: first, the resulting 4D gauge group
  is \E6, much larger than the SM gauge group, and secondly the number
  of \bs{27}'s is always larger than three.}.  Therefore, we work with
non-trivial solutions to ~\eqref{eq:ModInvV}, or {\it non-standard
  embedding} (NSE) models. On the string theory side, these correspond
to $(0,2)$ supersymmetric theories on the worldsheet. Moreover, also
from phenomenological requirements, we are interested in models with
non-trivial discrete Wilson lines. Indeed, all orbifold models with
promising phenomenological properties correspond to this
category~\cite{Saul}.  We describe below the massless spectrum of this
type of models.

\subsection{Massless orbifold spectrum}
\label{sec:masslessSpectrum}

The massless spectrum of an orbifold is comprised of two kinds of closed
strings. The {\it untwisted states} stem from the original spectrum of
closed strings of the heterotic string. In the supergravity limit,
they correspond to the components of the 10D supergravity multiplet and the
\E8\x\E8 vector multiplets that are left invariant by the orbifold
action.  The {\it twisted states} arise from closed strings attached
to the fixed points. These strings exist only due to the orbifold,
{\it i.e.}~they cannot be identified as part of the supergravity limit of
the heterotic string.  It is natural to associate each state with a
space group element commonly called {\it constructing element}.  {\it E.g.}~states in the untwisted sector are ``constructed'' by the element
$(\id,\,0)$ whereas those lying at the fixed point of the origin in
the $k=2$ twisted sector are associated with $(\vartheta^2,\,0)$.

The general form of an orbifold state associated with $g=(\vartheta^k,\,n_\alpha e_\alpha)$ can be written as
\begin{equation}
  \label{eq:gralState}
  \ket{\psi} = \ket{q_{\rm sh}}_R\otimes\beta\,\ket{p_{sh}}_L\otimes\ket{f}\,.
\end{equation}
In this formalism, $q_{\rm sh}=q+\bs{k}$ encodes the quantum numbers of the (right-moving) string in the
physical degrees of freedom of the compact and Minkowski space, where $q$ is an \SO8 weight 
({\it e.g.}~for $k=0$ $q$ is a weight of either
$\bs{8}_v$ or $\bs{8}_s$) and $\bs{k}=k\,v$ is the local twist vector defined above. $\beta$ denotes a
product of (left-moving) oscillator excitations $\widetilde\beta^{I}_{-1}$ in the 16 gauge dimensions,
$\widetilde\beta^{\mu}_{-1}$ in the Minkowski space, $\widetilde\beta^{i}_{-\eta_i}$ and
$\widetilde\beta^{\bar\imath}_{-1+\eta_i}$ in the holomorphic $i$ and antiholomorphic $\bar\imath$
directions of the compact space, respectively. Here $\eta_i=\bs{k}_i\mod 1$, such that $0<\eta_i \leq 1$. Further, $p_{\rm
  sh}=p+k\,V+n_\alpha\,{\cal A}_\alpha$ (with $p$ a weight of \E8\x\E8) is a 16D vector that carries the 
information on the gauge quantum numbers. Finally, the
component $\ket{f}$ contains the geometric information regarding the localisation in the compact space
of the orbifold state.

The quantum numbers of a massless state must fulfill
\begin{equation}
 \label{eq:Masslessness}
  \begin{array}{rl}
  \frac14 m_L^2 =& \frac12p^2_{\rm sh} + \widetilde N  - 1 + \delta c \stackrel{!}{=} 0\,,\\[2mm]
  \frac14 m_R^2 =& \frac12q^2_{\rm sh} - \frac12 + \delta c \stackrel{!}{=} 0\,,
  \end{array}
\end{equation}
where $\delta c = \frac12\sum_{i} \eta_i(1-\eta_i)$ corresponds to a change in the zero point energy due
to (twisted) oscillators. The oscillator number $\widetilde N$ is given by
\begin{equation}
  \label{eq:OscillatorNumber}
  \widetilde N= \sum_{i=1}^3 \eta_i \widetilde N^i_g + \bar\eta_i\widetilde N^{*i}_g
\end{equation}
with $\bar\eta_i=-\bs{k}_i\mod 1$, such that $0<\bar\eta_i \leq 1$.  Here, $\widetilde N^i_g$ and
$\widetilde N^{*i}_g$ count the number of excitations related to the constructing element $g$ 
in the $i$ and $\bar{\imath}$ directions.

Massless states $\ket{\psi}$ with constructing element $g=(\vartheta^k,\,n_\alpha e_\alpha)$
 acquire in general a non-trivial phase under the action of an 
arbitrary space group element $h=(\vartheta^l,\,m_\alpha e_\alpha)$,
\begin{equation}
 \label{eq:htrafo}
 \ket{\psi} \stackrel{h}{\longrightarrow} 
     e^{2\pi\I\,[p_\text{sh}\cdot (l V + m_\alpha {\cal A}_\alpha)- {\mathcal R} \cdot l v
    - \frac12\,\left((k V + n_\alpha {\cal A}_\alpha)\cdot (l V+m_\alpha {\cal A}_\alpha) - \bs{k}\cdot l v\right)]}\,\ket{\psi}\,,
\end{equation}
where we have defined the vector of {\it ${\mathcal R}$-charges}
 \begin{equation}
 \label{eq:Rcharges}
 {\mathcal R}=q_{\rm sh} - \widetilde{N}_g+\widetilde{N}^*_g\,.
 \end{equation}
The orbifold projection consists of letting space group elements $h$
such that $[g,h]=0$ act on the states 
$\ket{\psi}$. Only invariant states under this action are retained in the orbifold spectrum.

\subsubsection{Untwisted sector}

In the untwisted sector, $\bs{k}=0$ implies that the masslessness condition~\eqref{eq:Masslessness} reduces to
\begin{equation}
  \label{eq:MasslessnessUntwisted}
  \frac12p^2 + \widetilde N  - 1 = 0 = \frac12q^2-\frac12\,.
\end{equation}
Therefore, the solutions to this equation are given by $q^2=1$ and $p^2=2,\,\widetilde{N}=0$ or
$p^2=0,\,\widetilde{N}=1$. The resulting massless states can be gathered together in the following
categories:

\paragraph{Gravity multiplet and dilaton.} The states of the form $\ket{q}_R\otimes\widetilde\beta_{-1}^\mu\ket{0}_L$ 
($\mu$ runs in the Minkowski coordinates) such that
$q\cdot v=0\mod1$ correspond to a) the degrees of freedom of the graviton $g^{\mu\nu}$, b) the dilaton $\phi$, 
and c) one surviving component of the 10D $B$ 2-form, whose dual is
the so-called model independent axion, $a$. 
Since the orbifold preserves $\mathcal{N}=1$ supersymmetry, also the
superpartners are included in this notation.  We can build an
$\mathcal{N}=1$ chiral supermultiplet from the dilaton and axion,
whose bosonic component is given by 
\be
S = \sqrt{\det g} \, e^{-2\phi} + i a
\ee
with $g$ the internal metric.

\paragraph{Geometric moduli.} The geometric moduli in the effective theory are the pure internal
components of the 10D supergravity multiplet that survive the orbifold projection. In our notation, they are
expressed as $\ket{q}_R\otimes\widetilde\beta_{-1}^j\ket{0}_L$ and
$\ket{q}_R\otimes\widetilde\beta_{-1}^{\bar\jmath}\ket{0}_L$ with $(q-\widetilde{N}+\widetilde{N}^*)\cdot
v=0\mod1$. Traditionally, the states $\ket{q}$ have been denoted as $\ket{i}$ or $\ket{\bar\imath}$ 
depending on the position and sign of the vectorial \SO8 weight $q$. For instance, $\ket{1}$ denotes $q=(0,1,0,0)$  
and $\ket{\bar{2}}$ denotes $q=(0,0,-1,0)$. In this notation, states with $\ket{q}=\ket{i}$ and antiholomorphic 
oscillator index $\bar\jmath$ are the $h^{1,1}$ K\"ahler moduli. Since we will be working in  the  case with one K\"ahler modulus per 2-torus, then $i$ and $j$ coincide and the moduli are denoted as $T_i$.
States with $\ket{q}=\ket{i}$ and holomorphic oscillator
index $j$ are the $h^{2,1}$ complex structure moduli $U$. 

A more useful notation to illustrate the geometric origin of these moduli is achieved through the torus metric.
This  can be rewritten as $g_{\alpha\beta} =
e_\alpha\cdot e_\beta = R_\alpha R_\beta \cos \varphi_{\alpha\beta} 
$, with $R_\alpha$ being the ``radius'' of the cycle $e_\alpha$, and
$\varphi_{\alpha\beta}$ the angle between
$e_\alpha$ and $e_\beta$.  According to Eq.~\eqref{eq:Pong}, imposing $P$ invariance amounts to requiring
\begin{equation}
  \label{eq:gInvariance}
   \Theta^T g_{\alpha\beta} \Theta = g_{\alpha\beta}\,.
\end{equation}
This requirement fixes some of the parameters $R_\alpha$ and $\varphi_{\alpha\beta}$. The remaining free
parameters are combined  to build the $h^{2,1}$ K\"ahler moduli and $h^{1,1}$ complex structure
moduli. To this purpose, let us define the metric
of the $i^{th}$ $T^2_{(i)}$ plane as 
\begin{equation}
  \label{eq:2DMetric}
  g^{(i)}\equiv\left(
   \begin{array}{cc}
     g_{2i-1,\,2i-1} & g_{2i-1,\,2i}\\
     g_{2i-1,\,2i} & g_{2i,\,2i}\\
   \end{array}
  \right)\,.
\end{equation}
In terms of this metric, the complex structure and K\"ahler moduli of
the $i^{th}$ plane are given, respectively, by 
\begin{equation}
  \label{eq:UTDefinition}
  U_i = \frac{\sqrt{\det g^{(i)}}}{g^{(i)}_{11}} + \I\, \frac{g^{(i)}_{12}}{g^{(i)}_{11}}\,, \qquad
    \qquad T_i = \frac12 \left(\sqrt{\det g^{(i)}} + \I
    B_{2i-1,2i}\right)\,. 
\end{equation}
Since $U_i$ contain only ratios of terms in the metric, they describe
deformations in the
shape of the 2-torii.  Information about the overall size of the torii
is carried by $T_i$.  

\paragraph{4D gauge group.} The states of the form $\ket{q}_R\otimes\ket{p}$ with $q\cdot v=0\mod1$ and
$p\cdot V=0=p\cdot {\cal A}_\alpha$ together with the 16 Cartan generators of \E8\x\E8
$\ket{q}_R\otimes\widetilde\beta_{-1}^I\ket{0}_L$, $I=1,\ldots,16$, correspond to the gauge bosons of the
unbroken 4D gauge group $G_{4D}$. Notice that since $\beta_{-1}^I\ket{0}_L$
does not transform under the orbifold 
action, the rank of the gauge group is always preserved~\footnote{The
  rank of the gauge group can nonetheless be reduced {\it e.g.} by turning
on continuous Wilson lines~\cite{contWL}.}. However, out
of the 480 original charged gauge 
bosons ($p\neq0$) only a subset of them survives the orbifold projection. Therefore, the gauge group is
in general a subgroup of \E8\x\E8. For instance, SE induces the breaking $\E8\rightarrow\E6\x G_{2d}$,
with $G_{2d}=\SU3,\,\SU2\x\U1$ or $\U1^2$ depending on the \Z{N} orbifold. In non-standard embedding 
with Wilson lines, the breaking takes a more complex form.

\paragraph{Chiral multiplets.}

Untwisted charged matter arises from the states $\ket{q}_R\otimes\ket{p}$ where 
the orbifold projection is given by $p\cdot (l V + m_\alpha {\cal A}_\alpha)- q \cdot l v=0\mod 1$ 
with arbitrary integers $l$ and $m_\alpha$. Their gauge transformation 
properties with respect to $G_{4D}$ are encoded in $p$.

\subsubsection{Twisted sector}

Zero modes of the $\vartheta^k$ twisted sectors ($k\neq0$) are associated
to the constructing elements $g=(\vartheta^k,\,n_\alpha e_\alpha)$ and, therefore,
attached to the corresponding fixed points.  
These are matter states that take the form~\eqref{eq:gralState} and 
are constrained by the general masslessness condition~\eqref{eq:Masslessness}. 
As before, their gauge transformation properties are encoded in the gauge momenta $p_{\rm sh}$.
Note that, since at the fixed points the gauge symmetry can be larger than the 
one in 4D, from a local perspective, at the fixed points, the orbifold-invariant
states transform as complete multiplets of the local gauge symmetry. This observation 
has been useful to develop the concept of local GUTs~\cite{BHLR,Kobayashi:2004ya, Forste:2004ie}.

\bigskip

In the appendices we give the complete 4D massless spectrum for 
specific heterotic orbifold compactifications. For all models of
interest, the spectrum is composed of the supergravity multiplet and
the MSSM spectrum, together with some ``hidden'' gauge groups, 
hidden matter, 
a number of vector-like exotic matter multiplets and the bulk moduli.

\subsection{String selection rules}
\label{S:SelectionRules}

Couplings among matter fields are constrained by the symmetries of the
orbifold.  All invariant interactions with non-vanishing strength can
be identified by using the so-called 
selection rules~\cite{yukawas,Rule4}, which arise from those
symmetries. In this discussion~\footnote{The index $\alpha$, used here
to denote the various matter fields, should not be confused with the
lattice vector indices, $e_{\alpha}$.}, 
we consider the coupling
$c=\prod_\alpha A_\alpha$, where the matter fields 
$A_{\alpha}$ are characterised by the gauge  
momenta $p_{{\rm sh},\alpha}$, ${\mathcal R}$-charges ${\mathcal R}_\alpha$ and constructing element $g_\alpha$.

\paragraph{Gauge invariance.} Since the gauge quantum numbers of all matter states
are encoded in the gauge momenta $p_{\rm sh}$, $c$ is gauge invariant as long as it fulfills
\begin{equation}
\label{eq:gaugeinvariance}
 \sum_\alpha p_{{\rm sh},\alpha} = 0\,.
\end{equation}

\paragraph{${\mathcal R}$-charge conservation.} This selection rule of stringy origin arises from
the discrete symmetries of the 6D compactification lattice $\Lambda$ that are left unbroken 
under the orbifold action. In factorisable orbifolds, $c$ preserves these symmetries if it 
satisfies
\begin{equation}
\label{eq:Rchargeconservation}
\sum_\alpha {\mathcal R}_\alpha^i = -1\mod N_i\,,\qquad \text{ for all } i\,,
\end{equation}
where $N_i$ are the orders of the $T^2$ orbifold twists, {\it i.e.} $N_i$ are the smallest positive
integers such that $N_i v_i=0\mod1$ (no summation over $i$).

\paragraph{Space group selection rule.}
String interactions are only possible if the closed strings involved can join and form
other closed strings.
The constructing element $g_\alpha$ not only characterises the fixed point $f_\alpha$ of
the matter state $A_\alpha$ but also describes the boundary conditions of the associated
strings. Based on this, 
\begin{equation}
\label{eq:spacegroupsel}
 \prod_\alpha [g_\alpha] = (\id,\,0)\,,
\end{equation}
where $[g_\alpha]$ denotes some element of the conjugacy class of $g_\alpha$. The rotational part
of this equation reads $\prod_{k_\alpha}\vartheta^{k_\alpha}=\id$ or analogously
 $\sum_\alpha k_\alpha=0\mod N$, which is called {\it point group selection rule}.

For trilinear couplings, it is convenient to express the space group selection rule in terms
of the fixed points $f_{\alpha}$
\begin{equation}
  \label{eq:TrilinearSrule}
   (\id-\Theta^{k_{\alpha}})(\Theta^p f_\alpha-\tau_\alpha)+\Theta^{k_{\alpha}}(\id-\Theta^{k_{\beta}})(\Theta^q f_\beta-\tau_\beta)
    = (\id-\Theta^{k_{\alpha}+k_{\beta}})\Theta^r f_\gamma\,,
\end{equation}
where $\tau_{\alpha,\beta}$ are arbitrary lattice vectors and the integers $p,q,r$ lie between $0$
and $N/2$.

\bigskip

In the appendices, we give the Yukawa couplings allowed by the above selection rules for specific models. 
\section{Low Energy Effective Supergravity Theory}
\label{Sec3}

To study the dynamics of the moduli fields, we
construct the low energy effective field theory, which describes the
physics at 
energies well below the compactification scale.  
This is described by a 4D ${\mathcal N}=1$ supergravity theory and thus we must identify the
corresponding K\"ahler potential, $K$, gauge kinetic functions, $f_a$, and superpotential, $W$.

For the purely untwisted fields of the orbifold compactification, the action 
can be identified simply by directly truncating the low energy 10D
supergravity action describing the heterotic string.  Things are a bit
more complicated for twisted sectors, which emerge due to the
compactification.   The approach, pioneered in \cite{yukawas,DKL90},
is to construct an effective field theory that yields the same scattering
amplitudes as the full string theory does in the low energy limit.
Since orbifolds represent exactly solvable superconformal field
theories, all quantities can, in principle, be computed. 

The low energy effective action receives corrections from two
perturbative expansions: the string loop corrections, in 
$e^{2\phi}$, and
the supergravity approximation, in $\alpha'$.
 These corrections depend on the dilaton and volume moduli
 respectively. However, 
 4D ${\mathcal N}=1$ supersymmetry gives rise to powerful 
``non-renormalisation theorems''.  The tree-level superpotential cannot
receive corrections from (string loop or $\alpha'$) perturbative
effects, and is corrected only by non-perturbative effects \cite{WNR}.
Similarly, the gauge kinetic function receives string loop corrections
only at one-loop, and no higher \cite{fNR}. 

Finally, the low energy effective theory enjoys a further constraint from
the discrete target-space modular invariance, which is inherited from
the underlying conformal field theory.  This provides a valuable check on all computations.

In this section, we first describe in detail the modular symmetry, and
then present the terms of the low energy effective action for the
light fields in a generic orbifold compactification.  
We consider the untwisted moduli, $S,T_i, U_m$, gauge sector and matter
fields, $A_\alpha$.
When relevant, we distinguish between charged
matter fields (charged under the SM or hidden non-abelian groups), $Q_I \subset A_\alpha$, 
and SM and non-abelian singlets (charged only under hidden \U1s), 
$\Phi_{\Gamma}\subset A_\alpha$.

\subsection{Target space modular symmetry} \label{S:modinv}

The spectrum of states in an orbifold string compactification is invariant under certain 
discrete transformations on the moduli, together with the winding numbers and momenta.  
In the simplest orbifold compactifications, without Wilson lines, the corresponding
target space 
duality group is  
$SL(2,\mathbb{Z})_T^{h_{(1,1)}}
\times[SL(2,\mathbb{Z})]_U^{h_{(2,1)}}$. 
 Under this symmetry,
the moduli, $\phi_j = (T_i, U_m)$, and matter fields, $A_\alpha$,
transform as follows ~\cite{FLT,IL}
\bea
\label{eq:TModularTrafo}
  &&\phi_j\rightarrow \frac{a_j\phi_j-\I b_j}{\I c_j \phi_j + d_j}\qquad a_j,b_j,c_j,d_j\in \mathbbm{Z},\qquad a_jd_j-b_jc_j = 1 \\\nonumber\\
\label{eq:AModularTrafo}
&& A_{\alpha} \rightarrow M_{\alpha\beta}A_{\beta} \prod_{j}^{h^{1,2},h^{1,1}} (\I c_j \phi_j + d_j)^{n_{\alpha}^j}\, .
\eea
The modular weights $n^j_\alpha = (n_\alpha^i,\ell^m_\alpha)$ are given by~\footnote{In 
Eqs.~\eqref{eq:modularweights}, we have assumed that the moduli $T_i$ ($U_m$) lie in the 
$i^{th}$ ($m^{th}$) torus, which is the relevant case for us.}~\cite{IL}
\begin{subequations}
\label{eq:modularweights}
\begin{equation}
 \label{eq:modularweightsT}  
  n^i=\left\{
   \begin{array}{lcl}
    q_{{\rm sh},i}                             &\phantom{....} &\text{for } q_{{\rm sh},i}=0,-1\,,\\
    -1-q_{{\rm sh},i}+\widetilde{N}_g^i-\widetilde{N}_g^{*i} & &\text{for } q_{{\rm sh},i}\neq0,-1\,,
   \end{array}
  \right.
\end{equation}
and
\begin{equation}
 \label{eq:modularweightsU}  
  \ell^m=\left\{
   \begin{array}{lcl}
    q_{{\rm sh},m}                             &\phantom{....} &\text{for } q_{{\rm sh},m}=0,-1\,,\\
    -1-q_{{\rm sh},m}-\widetilde{N}_g^m+\widetilde{N}_g^{*m} & &\text{for } q_{{\rm sh},m}\neq0,-1\,,
   \end{array}
  \right.
\end{equation}
\end{subequations}
where the internal momenta $q_{{\rm sh}}$ and oscillator numbers
$\widetilde{N}_g,\widetilde{N}_g^{*}$ have been defined in
Section~\ref{sec:masslessSpectrum}.  The matrices $M_{\alpha\beta}$
are field-independent, and describe how the twisted fields with the
same weights and charges transform amongst each other.  Although in
the supergravity basis this transformation is not unitary, it is so
for the canonically normalised fields.

Together with the sigma-model symmetries, the target space modular
symmetry can be anomalous.  Part of this anomaly can   be cancelled
by a Green-Schwarz mechanism, which implies that the dilaton $S$
transforms at 1-loop, as \cite{anomalies}
\be
\label{eq:SModularTrafo}
S\rightarrow S - \frac{1}{8\pi^2}\sum_{j}^{h^{1,2},h^{1,1}} 
\delta_{GS}^j\log(\I c_j \p_j + d_j) \, ,
\ee
 where $\delta_{GS}^j$ are the gauge group independent Green-Schwarz
 coefficients describing 1-loop mixing between the dilaton and the
 geometric moduli.  The remainder of the anomaly is cancelled by
 threshold corrections due to massive string states, which we
 discuss below.

The K\"ahler potential, $K$, transforms at both tree and 1-loop level as 
\be
\label{eq:KModularTrafo}
K \rightarrow K + \sum_{j}^{h^{1,2},h^{1,1}}\log|\I c_j \p_j + d_j |^2 \, ,
\ee
\noindent and, in order to obtain invariant matter couplings, the 
superpotential has to transform (up to a field-independent phase) as \cite{LM} 
\be
\label{eq:WModularTrafo}
W \rightarrow \frac{W}{  \prod_{j}^{h^{1,2},h^{1,1}}(\I c_j \p_j + d_j)} \, .
\ee

Modular invariance means that several terms in the action can be
written in terms of modular forms, which transform covariantly with
some weight ${\rm k}$.  The most common of these is the well-known Dedekind eta function
\begin{equation}
  \label{eq:Dedekind}
  \eta(\phi) = e^{-\pi\phi/12}\prod_{n=1}^{\infty}(1-e^{-2n\pi\phi})\,,
\end{equation}
which is a cusp-form of weight ${\rm k}=1/2$.  This means that it
vanishes at real infinity~\footnote{We take the cusp for $SL(2,\mathbb{Z})$ at
real infinity.  Mathematicians instead take it at imaginary infinity.} and its modular 
transformation reads (up to a field-independent phase factor)
\begin{equation}
\label{eq:DedekindModularTrafo}
\eta(\phi)\rightarrow (\I c\phi + d)^{1/2} \, \eta(\phi)\,.  
\end{equation}

The presence of discrete Wilson line backgrounds breaks the modular
symmetries down to the subgroup that leaves the Wilson line invariant \cite{modWL,modWLBL}.
Typically, the surviving modular symmetries are congruence subgroups of $SL(2,\mathbb{Z})$,
which further restrict the integer parameters, $a, \, b, \, c, \, d$, of the transformations.
They can be computed by solving the constraints found in \cite{modWLBL}.  For instance, 
the groups that we encounter below are
\bea
&&\Gamma_0(N): c = 0 \,\, {\rm{mod}}\,\, N \cr
&&\Gamma^0(N): b = 0 \,\,{\rm{mod}} \,\, N \,\,\cr
&&\Gamma_1(N): a,d = 1\,\, {\rm{mod}}\,\, N \quad {\rm{and}}\,\, c = 0 \,\,{\rm{mod}}\,\, N  \,\, \cr
&&\Gamma^1(N): a,d = 1\,\, {\rm{mod}}\,\, N \quad {\rm{and}}\,\, b = 0 \,\,{\rm{mod}}\,\, N  
\eea
with of course also $a,b,c,d \in \mathbbm{Z}$ and $ad-bc = 1$.  It is easy to show~\footnote{Hint: 
write $N \frac{a \phi - \I b}{\I c\phi + d}$ as $\frac{a \, N \phi -\I N b}{\I (c/N) N \phi + d}$ 
and use (\ref{eq:DedekindModularTrafo}).} that, for instance, $\eta(N \phi)$ transforms covariantly 
under $\Gamma_0(N)$ but not under the full modular group $SL(2,\mathbb{Z})$.  For $\Gamma^0(N)$, 
the corresponding function is $\eta(\phi/N)$.

We close this subsection by noting that modular invariance represents
a powerful means to check the low energy effective theory that one
computes.  Moreover, identifying modular functions with the required
weights and zero/singularity structure provides a way to infer the moduli-dependence 
of various terms in the action, when direct computations are lacking.

\bigskip

In the remainder of this section, we introduce the different contributions to the low energy $\mathcal{N}=1$ 
4D effective theory below the compactification scale.  When reviewing previous computations, we present 
only the final results and refer the reader to the original literature for further details.

\subsection{Moduli and matter kinetic terms}

The kinetic terms for the moduli and matter fields are encoded in the
K\"ahler potential.   For the twisted fields they have been computed
to lowest order in the matter fields $A_\alpha$, with
\be
 | A_\alpha |^2 \prod_{j}^{h^{1,1},
   h^{2,1}}(\phi_j+\bar{\phi}_j)^{n_\alpha^j} \ll 1 \,.
\ee
Then, including one-loop effects, we have \cite{CCM,LM,BLST1,IL,DKL91}
\begin{eqnarray}
 \label{Kahler1}
  \hskip-0.1cm    K &=&    -\log\left[S+\bar S -
 \frac{1}{8\pi^2} \sum_j^{h^{1,1}, h^{2,1}} \delta_{GS}^j \log(\phi_j + \bar \phi_j)    \right] \nonumber \\
&& \hskip3cm  -\sum_j^{h^{1,1}, h^{2,1}}\log(\phi_j+\bar \phi_j)   
 + \, | A_\alpha |^2 \prod_{j}^{h^{1,1}, h^{2,1}}(\phi_j+\bar{\phi}_j)^{n_\alpha^j} \,.
\end{eqnarray}
Here $n_\alpha^j$ are the modular weights as defined before. 
It is easy to show that (\ref{Kahler1}) transforms in the required way
(\ref{eq:KModularTrafo}) under the modular symmetry.
The K\"ahler potential receives further perturbative corrections as
well as non-perturbative ones.  These, however, should be small
compared with the leading contribution above.

\subsection{Matter interactions}

At tree-level, there may be couplings between bulk fields, bulk fields
and twisted fields, or twisted fields located at the same fixed
points.  All these interactions can be understood within field theory,
and are of order one.  At the same time,
recalling the stringy nature of the twisted fields, even though their
center of masses are localised, they can stretch away from the
corresponding fixed points. 
This leads to non-perturbative corrections to the interactions between 
twisted fields, which can be interpreted as world-sheet instantons and
which are exponentially suppressed in the K\"ahler moduli that measure
the area of the wrapped instantons.

We now discuss in detail the trilinear  Yukawa couplings,
and their contributions to the superpotential.  Our focus here is on
twisted fields only, which have a non-trivial dependence of the K\"ahler moduli.  Then we briefly
comment on the aspects of higher order couplings that are relevant for
our purposes.

\subsubsection{Yukawa couplings}
The computation of trilinear Yukawa couplings was started in~\cite{yukawas, Burwick} and completed 
in~\cite{Erler:1992gt}. They were also explicitly computed for several orbifolds in~\cite{CGM}. 

Selection rules for permitted couplings were discussed in Section
\ref{S:SelectionRules}.  A generic coupling allowed between three
twisted fields from the sectors
$(\vartheta^k,\,\vartheta^l,\,\vartheta^{-(k+l)})$, located
respectively at
fixed points $(f_\alpha,\,f_\beta,\,f_\gamma)$, is
given by~\footnote{One can check that this expression is invariant under permutations of  $\alpha$, $\beta$, $\gamma$, as the physics
  dictates.  We have done so for our explicit couplings in the following section.},
\begin{equation}
\label{eq:Yuka1}
  \hskip-2cm Y^{kl}_{\alpha\beta\gamma} = F_{\alpha\beta\gamma} g_s\, \sqrt{V_\Lambda} \sum_u \prod_{i\notin {\mathcal J}} \sqrt{2\pi}
  \sqrt{\Gamma_{\bs{k}_i,\,\bs{l}_i}}
  \exp{\left(-\pi\, |\cot{\bs{k}_i\pi}+\cot{\bs{l}_i\pi}|^{-1}|u_i|^2\right)} \,,  
\end{equation}
where the local twists are $\bs{k}=k\,v$ and $\bs{l}=l\,v$, and
$\mathcal J$ is the set of indices $j$ labelling invariant planes, with
$\bs{k}_j, \,\bs{l}_j$ or $\bs{k}_j+\bs{l}_j = 0$ (mod 1).  The summation is over the components of
$u$ orthogonal to the planes $j\in {\mathcal J}$  and is due to contributions from fixed points equivalent to
$f_{\alpha,\beta,\gamma}$ thanks to lattice translations.
Furthermore, $F_{\alpha\beta\gamma}=\sqrt{l_\alpha l_\beta l_\gamma}$ 
is a normalisation factor, which counts the degeneracy due to fixed
points related to the $f_{\alpha}$ by rotations, with $l_\alpha$ being
the number of fixed points within the fundamental torus domain that
are related 
to $f_{\alpha}$ by powers of $\Theta$.  Also,
$V_\Lambda$ is the volume of the unit lattice orthogonal to the
invariant planes, with $V_\Lambda=\sqrt{\det{g}}$, and $|u|^2=u^t\,g\,u$.
Finally, 
\begin{eqnarray}
  \label{eq:latticeu}
  u_i &\in& [f_\beta-f_\alpha-\tau_\beta+\tau_\alpha +
  (\id-\Theta^{k+l})(\id-\Theta^{\gcd(k,\,l)})^{-1}\Lambda]_i\,, \\[1mm]  
  \label{eq:Gammas}
  \Gamma_{\bs{k}_i,\,\bs{l}_i} &=& \left\{
   \begin{array}{lcl}
    \frac{\Gamma(1-\bs{k}_i)\Gamma(1-\bs{l}_i)\Gamma(\bs{k}_i+\bs{l}_i)}{\Gamma(\bs{k}_i)\Gamma(\bs{l}_i)\Gamma(1-\bs{k}_i-\bs{l}_i)}\,,&\qquad& 0<\bs{k}_i+\bs{l}_i<1\,,\\
    \Gamma_{1-\bs{k}_i,\,1-\bs{l}_i}\,,&& 1<\bs{k}_i+\bs{l}_i<2\,,
   \end{array}
  \right.
\end{eqnarray}
where $\bs{k}_i\rightarrow\bs{k}_i\mod1$ so that $0<\bs{k}_i<1$, and $\tau_\alpha$ denotes vectors in
$\Lambda$ satisfying the space group constraint for non-vanishing
trilinear couplings~\footnote{Note that Eq.~\eqref{eq:TrilinearSruleForCouplings} differs
from~\eqref{eq:TrilinearSrule} by some factors $\Theta^p$. The reason is that the
rotational contributions to $Y_{\alpha\beta\gamma}^{kl}$ have been absorbed in the
coefficient $F_{\alpha\beta\gamma}$.}
\begin{equation}
  \label{eq:TrilinearSruleForCouplings}
  (\id-\Theta^k)(f_\alpha-\tau_\alpha)+\Theta^k(\id-\Theta^l)(f_\beta-\tau_\beta)
    =(\id-\Theta^{k+l}) f_\gamma\,.
\end{equation}

Let us briefly comment on how this result meets with our intuition on the couplings.  As one can 
see from the expression~\eqref{eq:Yuka1}, the coupling strengths depend on the orbifold, the fixed points and
the precise coupling considered. Moreover, the dependence on the $T_i$ moduli is also determined by the
particulars of the coupling, which are
encoded in the vector $|u|^2$ as well as in $V_\Lambda$. 
One can also see from this expression that, when the three twisted
fields are located at the same fixed point, the leading contribution
is perturbative  and of order  one.  Instead, when one or all three
fields lie on different fixed points, the coupling is exponentially
suppressed.
We calculate these couplings for concrete examples in Section~\ref{Sec4},
and observe all these features directly.

 We are now ready to construct a holomorphic superpotential from
the above Yukawa couplings.  For this we note that we have set
to zero possible backgrounds in the 
antisymmetric tensor, $B_{2i-1,2i}$, which provide the holomorphic
completion for the argument of the exponential in (\ref{eq:Yuka1}).  Recall 
also that 
Eq.~\eqref{eq:Yuka1} was computed for canonically normalised matter
fields.  The supergravity formula, relating these physical couplings
to the superpotential is \cite{wessbagger} (we suppress the $k,l$ indices)
\begin{equation}\label{Ytoh}
 Y_{\alpha\beta\gamma} = e^{\hat{K}/2}
\frac{h_{\alpha\beta\gamma}}{(K_{\alpha\bar\alpha} K_{\beta\bar\beta} K_{\gamma\bar\gamma})^{1/2}}\,, 
\end{equation}
where 
$\hat{K}$ is the tree-level K\"ahler potential for the (untwisted) moduli fields. 
Inverting this expression, we obtain the holomorphic superpotential
\begin{equation}
  \label{W1}
W_{\it yuk} = \sum_{\alpha,\beta,\gamma} h_{\alpha\beta\gamma}(T_i) A_\alpha A_\beta A_\gamma  \,.
\end{equation}

\bigskip

Finally we note that, according to the modular symmetry transformations for 
$A_\alpha$ (\ref{eq:AModularTrafo}) and $W$ (\ref{eq:WModularTrafo}), the Yukawa 
couplings must transform as \cite{LM,FLT}
\be
\label{eq:YukModularTrafo}
h_{\alpha\beta\gamma}(T_i) \rightarrow M^{-1}_{\alpha'\alpha}\,M^{-1}_{\beta'\beta}\,M^{-1}_{\gamma'\gamma}\,h_{\alpha'\beta'\gamma'}(T_i)
 \prod_{j}^{h^{1,1}}(\I c_j \p_j + d_j)^{-1-n^j_\alpha-n^j_\beta-n^j_\gamma} \,,
\ee
that is, the couplings transform, up to a non-trivial weight factor, amongst themselves for fields with the same modular weights and charges.

\bigskip

\subsubsection{Higher order couplings} \label{S:hots}

Higher order couplings are in principle also computable \cite{choiHO}.  Any allowed couplings of order $N$, will have the generic form in the
superpotential (in this section we put back explicitly the units)
\begin{equation}
  \label{Wint}
W_{\it int} =  M_{P}^{3}\, h_{\alpha_1\alpha_2\dots \alpha_N}(T_i) 
A_{\alpha_1} A_{\alpha_2} \dots A_{\alpha_N } \,. 
\end{equation}
where $h_{\alpha_1\alpha_2\dots\alpha_N}(T_i)$ has the required modular transformation properties and the fields $A_{\alpha_i}$ are dimensionless in Planck units. 

The higher order terms $N > 3$, are
non-renormalisable.  However, they may give rise to effective mass
terms if the coupling features $N-2$
non-Abelian singlet matter
fields, $\Phi_{\Gamma}$, which acquire a {\it vev} at some high scale.  In
this way, we may assume
that all SM exotics and
 charged hidden matter, $Q_I$, acquire an effective mass with
the mass matrix taking the form
\begin{equation}
  \label{eq:massMatrixEntry}
  \mathcal{M}_{IJ}=\sum_r M_{P} \, {h}_{{\Gamma}_{1}\ldots{\Gamma}_{r}
  IJ}(T_i) \, \langle {\Phi_\Gamma}_1 \dots {\Phi_\Gamma}_{r}\rangle\,. 
\end{equation}
This may be important to allow for gaugino condensation, discussed
below.   Assuming singlet {\it vevs} that lead to 
a universal mass scale for the 
SM exotics and charged hidden matter, the decoupling mass can be defined as 
\begin{equation}
  \label{eq:DecouplingScale}
  M_d = \det\mathcal{M}^{1/\dim\mathcal{M}}\,,
\end{equation}
 and $M_d \lesssim M_s \lesssim  M_{P}$ ($M_s = \alpha'^{-1/2}$ is the string scale) for singlet {\it vevs} slightly below $M_s$.
We discuss this in a little more detail at the end of this section.

\subsection{Gauge sector dynamics}

The final sector that we need is the gauge sector. We focus
on the hidden gauge group, $G = \prod_a G_a$.  The corresponding gauge couplings are field-dependent, and encoded
in the gauge kinetic functions.  In many cases, particularly when any
charged hidden matter is decoupled at some intermediate high scale,
$M_d \lesssim M_s$,
the gauge coupling becomes strong at some scale, $\Lambda<M_d$.  Below
$\Lambda$, the gauginos condense, leading to another (field
theoretical) non-perturbative contribution to the superpotential.

\subsubsection{Gauge kinetic function}
\label{S:f}
The tree-level gauge
kinetic function, $f_{tree} = k_a S$, with $k_a$ the level of the
Ka\v{c}-Moody  algebra, receives corrections at one-loop, due to
massive charged particles running in the loop:
\begin{equation}
  \label{eq:gaugekineticf}
   f_a~=k_a S + \Delta^{M_d}_a(T_i)  + \Delta^{M_s}_a(T_i, U_m) \,.
\end{equation}
Here, $\Delta^{M_d}_a(T_i)$ are the field theoretical threshold
corrections due to the massive charged hidden matter, and
$\Delta^{M_s}_a(T_i, U_m)$ are stringy threshold corrections due to massive
and winding string states.  The field theory contributions take the
form \cite{LM}
\begin{equation}
\label{eq:Thresholds}
\Delta^{M_d}_a(T_i) =- \frac{b_a - b_a^0}{8\pi^2} \,
\log\frac{M_d}{M_{P}}\,, 
\end{equation}
where the beta function coefficient,
\be\label{beta0}
b_a=-3C(G_a)+\sum_I T(Q^a_I)\,,  
\ee
includes the charged matter, whilst $b_a^0=-3C(G_a)$ is that of the pure Yang
Mills theory, valid below $M_d$.  Here, $C(G_a)$ and $T(Q^a_I)$ are the quadratic Casimirs in the
adjoint and matter representations, respectively. 

The stringy thresholds turn out to receive contributions only from
${\mathcal N}=2$ sectors in the string spectrum.  In the absence of Wilson lines, 
they were computed in~\footnote{There is generically an additional universal contribution
  to the thresholds, which also depends on the moduli
  \cite{Kaplunovsky:1995jw}, but has been little studied (see \cite{Omega}). 
We neglect it in the present work, but it would
be interesting to consider its effects in further studies.}
\cite{DKL91,Kaplunovsky:1995jw}, and can be written in terms of the Dedekind eta function as follows
\begin{equation}
\Delta_a(T_i, U_m) = -\frac{1}{8\pi^2}\sum_{j \in \mathcal J }
\frac{|P_j|}{|P|}\,(b_a^j)^{{\mathcal N}=2}\log{\eta(\phi_j)^2}\,.
\end{equation}
Here $j$ runs over all the complex structure and K\"ahler moduli associated to ${\mathcal N}=2$  planes. 
The $P_j$ refers to the subgroup of the point group $P$, which leaves unrotated 
the $j^{th}$ complex plane of $T^6$. Then $T^6/P_j$ is an orbifold with ${\mathcal N}=2$ 4D 
supersymmetry and $(b_a^j)^{{\mathcal N}=2}$ are the corresponding ${\mathcal N}=2$ beta function coefficients, given by
\be
  \label{eq:N2gralBetas}
  (b^j_a)^{{\mathcal N}=2} = -2 C(G_a) + \sum_{\alpha} T(R_\alpha^a)\,,
\ee
where $R_\alpha^a$ denote the representations w.r.t. the 
non-abelian group $G_a$, and
the summation runs over all half-hypermutiplets of the theory.
We discuss the modifications of these formulae due to the
presence of Wilson lines in the following section.

Notice that, since the Dedekind function vanishes at infinity, the
threshold corrections diverge there.  This divergence is due to the
fact the Kaluza-Klein modes all become massless as the torus volume
goes to infinity, signalling a breakdown of the 4D effective theory.  The
modular symmetry ensures there is a similar divergence as the
torus volume goes to zero, when the winding modes become massless.  If
there are special points in the moduli space where extra charged modes become
massless, we expect additional, milder singularities in the thresholds
at those points.  

Recalling that the stringy threshold corrections help to cancel the
modular and $\sigma$-model anomalies, we can relate their coefficients
to those of the anomalies, $b_a^{'j}$, in the following way \cite{DKL91,Kaplunovsky:1995jw}
\begin{equation}
 \label{eq:relationbeta-delta}
  \frac{|P_j|}{|P|}\,(b_a^j)^{{\mathcal N}=2} = b_a^{'j}-k_a \delta_{GS}^j\,.
\end{equation}
The anomaly coefficients are computed to be
\begin{equation}
  \label{eq:modularanomaly}
   b_a^{'j} = -C(G_a) + \sum_{I} T(Q^a_I)(1 + 2 n_{I}^j) \,
\end{equation}
and so the relation (\ref{eq:relationbeta-delta}) provides a way to calculate 
$\delta_{GS}^j$. 

Having established the holomorphic gauge kinetic function, the expression for the running, loop corrected gauge
coupling constant $g_a$ at a scale $\mu$  is \cite{LM} 
\begin{eqnarray}
\label{ga}
\frac{1}{g_a^2(\mu)} =&& \frac{1}{g_s^2}
+\frac{b_a^0}{16\pi^2}\,\log{\left(  \frac{M^2_d}{\mu^2}   \right)} +
\frac{b_a}{16\pi^2} \, \log{\left(\frac{M_s^2}{M_d^2}\right)} \cr
&& -
\frac{1}{16\pi^2} \sum_{j \in \mathcal J} \left(\frac{b_a^0}{3}-k_a \delta_{GS}^j\right)\log{(\phi_j
      + \bar \phi_j)} \cr
&&     - \frac{1}{16\pi^2} \sum_{j \in \mathcal J} \left(b_a^{'j}-k_a \delta_{GS}^j\right)\log{|\eta(\phi_j)|^4} \,,
\end{eqnarray}
where the gauge coupling at the string scale is given by 
\begin{equation}
\frac{2}{g_s^2} =  S + \bar S -\frac{1}{8\pi^2} 
 \sum_{j \in \mathcal J }\delta_{GS}^j \log{(\phi_j + \bar \phi_j)} \,, 
\end{equation}
and again, the sum over $j$ runs over $h^{2,1}$ moduli and K\"ahler moduli associated to 
${\mathcal N}=2$ planes.  Notice that, whereas in the supergravity holomorphic gauge 
kinetic function the threshold corrections kick in at the fundamental Planck scale, 
for the real gauge couplings $g_a$ the relevant scale is the string scale.  Moreover, the gauge
couplings must not only be real, but also modular invariant.  Modular invariance 
is guaranteed by the effect of
massless gauginos running in the loop, which give rise to the fourth
term in \eqref{ga}.  For an enlightening discussion on holomorphic gauge
kinetic functions and modular invariant gauge couplings, see~\cite{louishol}.

\subsubsection{Gaugino condensation}
\label{S:Wgc}
In the effective theory below the decoupling scale, $M_d$, we are left
with pure Yang-Mills groups.  In this case, the gauge couplings
generically become strong at some scale $\Lambda_a$ and the following
superpotential is generated 
\begin{equation}
\label{eq:condensateW}
  W_{\it gc} \approx  \sum_a \, d_{a}\, \exp\left(\frac{24\pi^2}{b_a^0} f_a \right)\,,  
\end{equation}
where $d_{a}$ is a constant that arises in the process of integrating out the condensate, and is given, up to an unknown constant $c$, by
\be
\label{da}
d_a = \frac{c}{e} \frac{b_a^0}{96 \pi^2}
\ee
in Planck units (see {\it e.g.} \cite{BLreview}) and $e$ is just the Euler number.

It is a simple exercise to show how
$W_{gc}$ transforms under modular transformations
(\ref{eq:WModularTrafo}).  Using the above expressions for $f_a$ (\ref{eq:gaugekineticf})-(\ref{eq:modularanomaly}), the
definition of $M_d$ (\ref{eq:DecouplingScale}) and the modular transformation of $\det\mathcal{M}$ that follows from~\eqref{eq:AModularTrafo}
and~\eqref{eq:YukModularTrafo},
\begin{equation}
  \label{eq:detMModularTrafo}
  \det\mathcal{M}\rightarrow \det\mathcal{M} \prod_{j}^{h^{1,2},h^{1,1}}(\I c_j \p_j +
  d_j)^{-\sum_I(1+ 2 n^j_I)}\,,
\end{equation}
one can  verify that $W_{gc}$ transforms in the required way
(\ref{eq:WModularTrafo}), provided that the charged hidden matter is
such that
\begin{equation}\label{matconst}
 \sum_I T(Q_I) \sum_J \left[(1+2 n^j_J)\right] - \dim\mathcal{M}
 \sum_J \left[(1+2 n_J^j) T(Q_J)\right] = 0 \, .
\end{equation}
This appears to be a stringy constraint on the matter content, which
explicit models -- consistent with all our assumptions -- should fulfill.
In the models analysed in Section \ref{Sec4}, we show that this non-trivial condition is indeed satisfied.

\bigskip

\subsection{The scalar potential}

With the K\"ahler potential, $K$, superpotential, $W$, and gauge
kinetic functions, $f_a$, in hand, we can now compute the scalar
potential of the action.  There are two kinds of contributions:
\be
V= V_F + V_D\,.
\ee
The F-term potential is given by
\be
\label{scalarV}
V_F= e^K \left(K^{A\bar{B}}D_AW {D_{\bar B}\bar{W}} - 3|W|^2 \right)\,,
\ee
where the indices $A,B$ label all the chiral supermultiplets present and $D_A W = \partial_A W +  \partial_A K  \, W$. 
For our generic K\"ahler potential, Eq.~\eqref{Kahler1}, and a general
$W$, (\ref{scalarV}) takes the form
\bea\label{FtermV}
&& V_F= e^K \Bigg[   \left| Y W_S -W \right|^2 
+ \sum_\alpha\frac{1}{\prod_j  (\phi_j + \bar {\phi_j})^{n^j_\alpha}} 
\left| W_{A_\alpha} + \overline{A}_\alpha
\prod_i  (\phi_i + \bar {\phi_i})^{n^i_\alpha} W \right|^2  \nn \\
&& \hskip1cm  + 
\sum_i \frac{Y}{Y - \delta^i - Y \sum_\alpha |A_\alpha|^2 \,n^i_\alpha \prod_j  (\phi_j + \bar {\phi_j})^{n^j_\alpha}} \times \nn \\
&&\hskip2.5cm 
\bigg| \delta^i \, W_S +  (\phi_i + \bar {\phi_i}) W_{\phi_i} 
- W 
- \sum_\alpha n_\alpha^i  A_\alpha W_{A_\alpha}  \bigg|^2 - 3|W|^2 \,\,  \Bigg]\,,
\eea 
where we have defined
 $Y = S + \bar S -\sum_j \delta^j \log{( \phi_j +\bar \phi_j )}$ and 
 $\delta^j = \frac{\delta_{GS}^j}{8\pi^2}$. 
 
 Additionally, there are $D$-term contributions to the scalar potential
 from each gauge multiplet, abelian and non-abelian alike.  In
 general, these can be written as
\be 
  V_{D_a} = -\frac12 ({\rm Re}f_a)^{-1} \left(K_{\alpha} t^{\alpha\beta}_{a} A_{\beta}\right)^2\,,
\ee 
where $t^{\alpha\beta}_{a}$ are the generators of the gauge
 group, appropriate to the representation in which the chiral
 superfields $A_{\alpha}$ lie.  However, amongst all the gauge
 multiplets, there is one special one, which is associated to a
 pseudo-anomalous $\U1_X$ gauge symmetry.  The would-be anomaly is
 cancelled by a Green-Schwarz mechanism, which requires the dilaton
 $S$ to transform non-linearly under the $\U1_X$ at one-loop, and
 which is associated to a field-dependent Fayet-Iliopoulos term.  In
 this case, the contribution to the $D$-term potential is
\be
V_{D_X} = -\frac12 ({\rm Re}f_X)^{-1}\left(K_{\alpha} q_{\alpha} A_{\alpha} + \frac{\delta_{GS}}{Y}  \right)^2\,.
\ee 
Here, $q_{\alpha}$ are the $\U1_X$ charges, and $\delta_{GS}$ the $\U1_X$ Green-Schwarz anomaly 
coefficient, which is given by \cite{deltaGSu1}
\be
\delta_{GS} = \frac{1}{96 \pi^2} {\rm tr} \,  t_X\,, 
\ee
where $t_X$ is the generator of $U(1)_X$.
 An important remark is in order here. It is clear
from the expression above that all geometric moduli appear in the
D-term potential via $K_\alpha$, unless the modular weights vanish.
In orbifold models, matter fields always have non-vanishing modular weights in at least one torus, thus this 
dependence is always present. 
Moreover, a non-trivial moduli dependency is also ensured via the perturbative correction to the dilaton appearing in $Y$.

\subsection{Decoupling of exotics} \label{S:exotics}

To complete this section, we make a few more remarks on the decoupling
of exotics, first introduced in Subsection \ref{S:hots}.  Recall that
the orbifold compactification initially gives rise to a spectrum which
includes not only the particles of the MSSM, hidden gauge group and moduli, but also a
number of vector-like exotic matter and SM singlets, which may be
charged under non-Abelian hidden sectors or only under hidden \U1's.  
These must all be decoupled somehow if we
are to describe Nature.  

One way that this may happen is if
non-trivial {\it vevs} are induced for some
non-Abelian singlets, thanks to {\it e.g.} the Fayet-Iliopoulos term
described above.  As explained in Subsection \ref{S:hots}, the
singlet {\it vevs} subsequently give masses to the exotics and charged
hidden matter, due
to their higher order couplings.  At the same time, the hidden \U1
gauge symmetries will typically all be broken.  This mechanism has
been studied in \cite{Saul}, where it was argued that the {\it vevs}
can be turned on in a way that preserves supersymmetry, along the
lines of \cite{mononomials}.  Important to
our picture, is that the {\it vevs} induced by the D-terms can be arranged
to be constant with respect to the moduli and that the coupling
strengths that give rise to effective mass terms are also moduli
independent.  Indeed, otherwise, we would not have effective mass
terms but couplings between the exotics and the light scalar moduli
fields.
Although such moduli independence is not generic, it does seem to be
possible, and, moreover, an interesting way to allow for an
additional scale in the problem.  It would
be important to investigate these issues further 
\footnote{The possibility to dynamically decouple 
matter fields via FI D-terms in our explicit models turns out to be 
highly involved. Note that, as mentioned before, the FI D-terms have 
a non-trivial dependence on all the moduli, via their modular weights 
and 1-loop corrections. This possibility certainly deserves further 
investigation.}.

For the time being, we note some significant phenomenological
motivations for our 
introduction of an additional scale.  On one hand we would like the exotic
matter to be decoupled at least at the GUT scale, in order to maintain
the gauge coupling unification of the MSSM.  On the other hand, we
would like the supersymmetry breaking scale, usually associated with
the moduli stabilisation, to be down at the electroweak scale in order
to understand the latter~\footnote{One alternative (but less motivated) picture is that the
  moduli are somehow stabilised at a high scale, the exotics are then
  decoupled via the singlet {\it vevs} somewhere below this scale as in
\cite{Saul}, and
  supersymmetry is broken later by some so far unknown mechanism.  As
  yet another possibility, albeit technically unfeasible, all the
  fields may be fixed at the same time. FI D-terms~\cite{dterm} and matter fields active during
  moduli stabilisation~\cite{dterm, matterdS} have been noted to help produce
  de Sitter/Minkowski   
vacua.}.  Furthermore, the decoupling
of charged 
hidden matter at some high scale can be important to ensure
that gauginos condense later, since massless charged matter would drive
the ${\mathcal N}$=1 beta-function coefficients towards positive
values which destroy asymptotic freedom.  This especially applies to
gauge groups of small rank, as they tend to come with a lot of hidden
matter in heterotic orbifold compactifications.  Indeed it would
otherwise be very difficult to find multiple condensing gauge groups
in the minilandscape.  Finally, it should be remembered that gaugino
condensation in the presence of 
massless charged hidden matter is actually poorly understood (see
{\it e.g.} \cite{CCM}). 

Therefore, in the following, we simply assume that all the exotics are
decoupled below some scale near the string scale, $M_d$, consistently with supersymmetry.  In our low energy 4D effective 
supergravity theory the physics of the decoupling is then
parameterised by the singlet {\it vevs} and the scale $M_d$.
Indeed, since supersymmetry is preserved, we can simply integrate out
the heavy matter by replacing them with their {\it vevs} in the K\"ahler
potential, superpotential and gauge kinetic functions.  
The action that results then describes the MSSM sector
and the bulk moduli, and (since the ``hidden" \U1's are all broken) 
the relevant contributions to the scalar potential can only be from F-terms.

\section{Moduli Stabilisation in Realistic \bs{\Z6}--II Orbifold Models}
\label{Sec4}

We are now ready to study the dynamics of the moduli that results from
the action described above.  In this section, we compute explicitly
all the supergravity ingredients that we reviewed in previous
sections, in concrete orbifold models. We stress here that whilst most
of the ingredients have been discussed in the past, this is the first
time that such results are applied
 to ``realistic'' orbifold examples
in their full glory.  Perhaps unsurprisingly, we find that several of
the features that have been used in the toy-models constructed to
date, tend not to be observed in real models.  At the same
time, we will see that whilst the scalar potential seems to have
sufficient structure to stabilise all of the bulk moduli, we found 
only unstable vacua.  Whether or not these
instabilities are inevitable is currently under investigation \cite{yashar}.
Meanwhile, since all the fields typically contribute to the breaking
of supersymmetry, we generically obtain de Sitter vacua.

We start by  making some  general observations and then
proceed by analysing explicit models in which the exact MSSM can be found.

\subsection{Moduli stabilising contributions}

Let us first explain how the ingredients discussed in the previous
section give rise to 
a scalar potential for all the untwisted moduli.

As described above, the Yukawa couplings between three twisted fields
receive world-sheet non-perturbative instanton contributions, which
depend on the area of the cycles that the instantons wrap.  This gives
rise to a potential for the K\"ahler moduli of the corresponding
cycles.  Notice that if any of the twisted sectors involved has an
invariant plane, then the corresponding twisted fields are not
localised in that plane, so there is no area suppression in the
couplings there. 
Therefore, $W_{\it yuk}$ depends only on the K\"ahler moduli 
belonging to planes that are rotated by the twists involved.  

Meanwhile, it is well-known that gaugino condensates in hidden
gauge groups give rise to a non-trivial potential for the dilaton,
$S$.  Moreover,  
the threshold corrections to the gauge couplings lead to a dependence
in $W_{\it gc}$ on all the moduli
describing 2-torii that remain invariant
under one of the orbifold twists.  In this way, the gaugino
condensates provide a non-trivial potential for all the complex
structure moduli and some of the K\"ahler moduli, in particular all those that did not appear
in the Yukawa couplings. 
Therefore, in principle, none of the $S, T_i, U_m$ are actually flat
directions.  We now ask, in  concrete models, if it is possible to
stabilise all moduli and to obtain a de Sitter vacuum.

\subsection[Explicit \Z6--II MSSM models]{Explicit \bs{\Z6}--II MSSM models}
\label{sec:ConcreteExample}
In this section we consider two representative MSSM models from the
\Z6--II  minilandscape~\cite{Saul}, which exhibit however different
generic  properties for the moduli dynamics. They   
serve as good examples of the sort of structure one expects to see in
the fertile heterotic orbifolds~\footnote{Although the models presented 
here are based on \E8\x\E8 heterotic \Z6-II orbifolds, similar constructions 
exist in the context of its \SO{32} 
sister and other \Z{N} 
orbifolds~\cite{RamosSanchez:2008tn}.}.
We now introduce the essential features of the models with the details left for the appendices.  

 \FIGURE[t!]{\epsfig{file=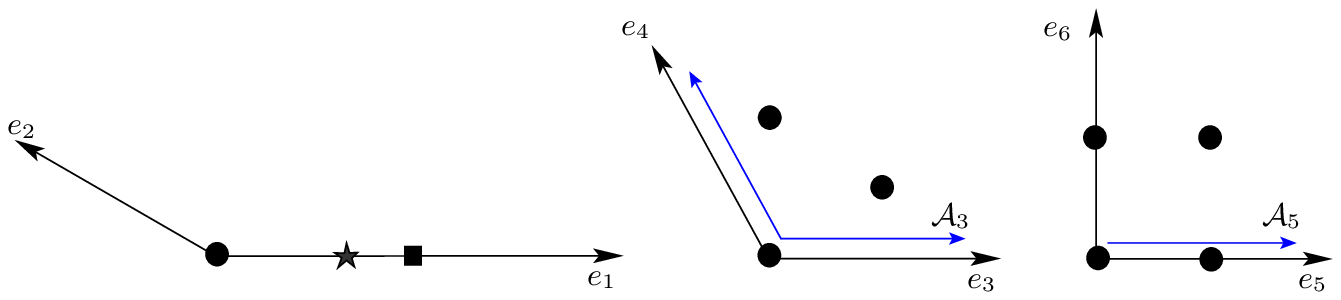}
 \caption{Geometry of the $T^6$ torus of a \Z6--II orbifold. The big dots denote points left invariant by
  $\vartheta$, the points denoted by a star and a square are fixed only under $\vartheta^2$ and
  $\vartheta^3$ respectively. In addition, the shorter/blue arrows denote 
  the non-trivial Wilson lines given in Appendices~A and B.}
  \label{fig:Z6IIGeometry}}
The twist for the \Z6--II orbifold is $v =\frac{1}{6} (0,\,1,\,2,\,-3)$, 
for which an admissible choice of $\Lambda$ is the root lattice of $G_2\x\SU3\x\SO4$ depicted in
Figure~\ref{fig:Z6IIGeometry}. 
The action of the corresponding Coxeter element on the lattice basis is given by  
\bea
&& \Theta e_1 = 2e_1 + 3 e_2 \,, \qquad \Theta e_2 = -e_1 - e_2\,, \nn \\
&& \Theta e_3 = e_4 \,, \qquad\quad  \qquad \Theta e_4 = -e_3 - e_4\,, \nn\\
&& \Theta e_5 =  - e_5 \,, \,\,\qquad \qquad \Theta e_6 = -e_6\,. 
\eea

Point group invariance of the metric, Eq.~\eqref{eq:gInvariance}, implies that there are five free parameters 
(associated with three real K\"ahler moduli and one complex structure modulus). 
In total we have the following relations among the parameters (in string units, {\it i.e.} $R_i^2\equiv R^2_i/\alpha'$)
\begin{equation}
R_1 = \sqrt{3} \, R_2\,, \qquad R_3 = R_4\,, \qquad \cos\varphi_{12} = -\frac{\sqrt{3}}{2}\,,
\qquad \cos\varphi_{34} = -\frac{1}{2}   
\end{equation}
and the free parameters are $R_1, \, R_3,\, R_5,\, R_6,\,\cos\varphi_{56}$.
Subsequently, using the standard definition of K\"ahler and complex 
structure moduli, Eq.~\eqref{eq:UTDefinition}, we find that the
geometric parts of our moduli are given by 
\begin{eqnarray}
&Re(T_1) \equiv t_1  =  \frac{R_1^2}{4\sqrt{3}} \,, \quad 
Re(T_2) \equiv t_2  =  \frac{\sqrt{3}}{4}  \,R_3^2  \,, \quad 
Re(T_3) \equiv t_3  = \frac12 R_5\,R_6 \,\sin{\varphi_{56}} \,,& \nonumber\\
\nn \\ 
\label{eq:ModelModuli}
&U_3 \equiv u_3 + \I\nu_3   =  \frac{R_6}{R_5}\, \sin{\varphi_{56}} + \I \,\frac{R_6}{R_5}\, \cos{\varphi_{56}}\,,&
\end{eqnarray}
where $\varphi_{56}$ is the angle between $e_5$ and $e_6$.

The different models are obtained by choosing different gauge embeddings ({\it i.e.}~different shift vectors and Wilson lines).
We now present each in turn, and write down explicitly their low
energy effective theories by applying the results of Section 3.  We
assume for simplicity that non-Abelian singlets acquire {\it vevs} as
discussed in Subsection \ref{S:exotics}, which are
constant $A_\alpha=\langle A_\alpha \rangle$ and of about
the same order.  Moreover, we assume that all the 
exotics and charged hidden matter are subsequently decoupled at the same scale, $M_d$.  
We then focus on the moduli and hidden gauge group only,
since the observable matter must 
have vanishing expectation values in order to preserve the SM gauge
group.  Our dynamical fields are thus $S, T_i, U_3$, $i=1,2,3$.

\subsubsection{Model I: double gaugino condensate}

The shift vector and Wilson lines for this model are given in 
Appendix~\ref{sec:Model1}.  The unbroken gauge group after compactification 
is $ \SU{3}\x\SU{2}\x \U1_Y \x\left[\SO{8}\x\SU{3}\x\U1^6\right]$. The full 
4D massless spectrum is given in Appendix ~\ref{app:4Dspectrum}.
As explained above, the presence of discrete Wilson lines breaks the
modular symmetry down to a congruence subgroup of $SL(2, \mathbb{Z})$. 
In the present model, we compute the modular symmetry group to be
$SL(2,\mathbb{Z}) \times \Gamma_1(3) \times \Gamma_0(4) \times
\Gamma^0(4)$, where the transformations act on $T_1,T_2,T_3$ and $U$
respectively.  We give some useful information about these groups in
Appendix \ref{app:modsym}.
The resulting action is  as follows.

\paragraph{K\"ahler potential.}

The K\"ahler potential is given by (\ref{Kahler1}), where the modular
weights can be found in Table \ref{tab:4DSpectrumModel} and the
Green-Schwarz coefficients are computed in Appendix
\ref{app:deltaGS}. They are
\begin{equation}
  \label{deltaGSs}
  \delta_{GS}^1 = -\tfrac{19}{3}\,,\quad \delta_{GS}^2 =-\tfrac73\,,\quad \delta_{GS}^3 =0\,.
\end{equation}
In principle, all the SM singlet {\it vevs} contribute to $K$. Since they must 
be small in order for us to maintain perturbative control, we can safely neglect them at this level.

\paragraph{Gauge kinetic function.}

We can compute the gauge kinetic function by referring to Section \ref{S:f}.  
The values for the beta-function coefficients that measure the field
theoretic threshold corrections 
turn out to be~\footnote{Here we used $C(SO(N))=N-2$, $T({\rm\bf N}_{SO(N)})=1$,
  $C(SU(N))=N$, $T({\rm\bf N}_{SU(N)})=\frac12$.}
\bea
&& b_{SO(8)} = -9, \qquad \, b^0_{SO(8)} = -18 \cr
&& b_{SU(3)} = 1,  \qquad \quad b^0_{SU(3)} = -9 \, .
\eea
Meanwhile, although stringy threshold corrections in the presence of
discrete Wilson lines have been scarcely studied~\footnote{See the
  second reference of \cite{modWL}, and very recently \cite{ThresWL}.},
we can infer their modular dependence from the modular symmetries and
expected singularity structure.   For example, assuming that the
thresholds depend on $T_2$ as $\log [\eta(3\, T_2)^2]$ leads to the required
covariance under the modular symmetry $\Gamma_1(3)$ in the second
plane, and reproduces the expected divergences as the volume of that
plane goes to zero or infinity (see the paragraphs below Eq. (\ref{eq:N2gralBetas}))~\footnote{We do not exclude other
  dependencies, for instance one that is covariant under $\Gamma_1(3)$
but not $\Gamma_0(3)$.  Indeed, direct computations of the stringy thresholds in
the presence of Wilson lines would be very valuable.}.  Moreover, as in the case without Wilson lines,
we expect the contributing sectors to be the ${\mathcal{N}}=2$
ones, and this allows us to compute the coefficients of the
thresholds.  We study the corresponding auxiliary ${\mathcal{N}}=2$ theories in Appendix
\ref{app:N2theories}.  

Putting all information together, the final result for the two hidden gauge
groups, $SO(8)$ and $SU(3)$, are respectively
\bea
f_8 &=& S - \frac{9}{8\pi^2} \log{\frac{M_d}{M_{P}}} +  \frac{2}{3\pi^2}
 \log\eta\left(3T_2\right)^2 - \frac{1}{8\pi^2} \left(
 \log\eta\left(4T_3 \right)^2  + \log\eta\left(U/4\right)^2\right) \, ,\cr
 \nn \\
f_3 &=& S - \frac{5}{4\pi^2} \log{\frac{M_d}{M_{P}}} -  \frac{1}{12\pi^2}
 \log\eta\left(3T_2\right)^2 - \frac{3}{8\pi^2}
 \left(\log\eta\left(4T_3 \right)^2  + \log\eta\left(U/4\right)^2 \right)\,. 
\eea

At this point one can check  explicitly, using the data in the
Appendix \ref{app:4Dspectrum}, that the constraint (\ref{matconst}) is
indeed satisfied, and the expressions are modular covariant.  Another
nice check of the threshold coefficients is to note that computations for all the gauge
groups lead to the same Green-Schwarz anomaly coefficients for the
sigma-model and modular anomalies, via Eq. (\ref{eq:relationbeta-delta}).

\paragraph{Superpotential.}

The superpotential for the moduli receives contributions
from Yukawa couplings, higher order matter couplings and gaugino
condensation.  In the moduli potential, we neglect the higher order couplings, since they will
be suppressed with respect to the trilinear Yukawa couplings by the Planck scale
and the small matter {\it vevs} (see however \cite{cvetichots}). Also,
we consider only couplings between (non-Abelian) singlets, since we allow
non-trivial {\it vevs} only for those fields.

The allowed Yukawa couplings are computed in
 Appendix~\ref{app:3pointCouplings}.  All those between gauge
singlets are of the kind $\vartheta^2\vartheta^5\vartheta^5$.  
From (\ref{eq:Yuka1})-(\ref{eq:TrilinearSrule}), we can explicitly compute
this coupling in terms of the $T_i$ moduli (see Appendix \ref{app:Yukawas} for details).  The result is:
\be\label{Y25}
 Y^{25}_{\alpha\beta\gamma} = F_{\alpha\beta\gamma}\,g_s \sqrt{t_1 t_2} \, N_{255} \sum_{u_1,u_2}\exp{(-{\it b}_{u_1}\, T_1 - {\it b}_{u_2} \,T_2)}\,, 
\ee
where
\begin{eqnarray}
  \label{eq:N345N255}
&&    N_{255}=\frac{2^{14/3}\pi^4}{3^{3/2}\Gamma^6(1/3)} 
\,,\\ \nn \\
  \label{eq:345255Exponentials}
 &&  {\it b}_{u_1}= \pi u_1^t  \, L_1 \, u_1\,, \qquad
   {\it b}_{u_2}= \pi u_2^t \,L_2 \, u_2\,,\\ \nn \\ 
&&
\label{eq:ML}
 L_1 = \left( 
     \begin{array}{ccc}  
         6 && -3 \\
         -3 && 2   
     \end{array} \right)\,, \qquad   \qquad
 L_2 = \left( 
     \begin{array}{ccc}
         2  && -1   \\
         -1 && 2   
      \end{array} \right) \,.
\end{eqnarray}
The 2D lattice vectors $u_i$ depend on the fixed points $(f_\alpha,\,f_\beta,\,f_\gamma)$ of the
fields involved in the coupling, and are given
in~\ref{app:Yukawas255}, as are the normalisation factors
$F_{\alpha\beta\gamma}$. 

The corresponding contributions to the holomorphic superpotential are
computed using (\ref{Y25}) and (\ref{Ytoh}).  In Appendix
\ref{app:modinvyuk} we explain the necessary tricks to show that the
explicit expressions are modular covariant, as they must be.  Here, we
consider only 
the leading order terms in the sum 
over instantons, and replace the SM singlets by their {\it vevs}
$A_\alpha$.  The result is~\footnote{We have taken the prototype couplings 
$n_9 n_{52} n_{64}$,   
$n_{23} n_{52} n_{64}$, $n_{9} n_{59} n_{66}$,  $n_{23} n_{59}
n_{66}$ (see Table \ref{tab:3pointCouplings}),  and assumed
consistently, for simplicity, the {\it vevs}  $n_{52} = n_{64} = n_{59} = n_{66} = A_1$, $n_9
= A_2$, $\sqrt2 n_{23} = A_3$.} 
\begin{equation}
  \label{eq:superpotentialPart1}
  W_{\it yuk} = 2 \, N_{255}\, A_1^2 \, e^{-2\pi \, T_2/3} \left(A_2 + A_3 e^{-2\pi \, T_1/3}\right) \,.
\end{equation}

Finally, with the gauge kinetic functions in hand, we can immediately write down the contribution to the superpotential that arises when the gauginos in the hidden sector
condense:
\bea\label{Wmodel1}
&& W_{\it gc} = -\frac{c}{e}\,\frac{3}{16 \pi^2} \,\, e^{-4\pi^2 S/3} \, \left(\frac{M_d}{M_P}\right)^{3/2}\,        
    \eta(3\,T_2)^{-16/9} \, \left[ \eta(4\,T_3)\,\eta(U/4)\right]^{1/3}\nn \\
 && \hskip 2cm
  -\frac{c}{e}\,\frac{3}{32 \pi^2} \, e^{-8\pi^2 S/3} \, \left(\frac{M_d}{M_P}\right)^{10/3}\,        
    \eta(3\,T_2)^{4/9} \,\,  \left[ \eta(4\,T_3)\,\eta(U/4)\right]^{2} \,,
\eea
where we have used the definition of $d_a$  in (\ref{da}).  Notice
here that several of the Dedekind eta functions, originating from the
stringy threshold corrections, appear with positive
powers, in contrast to what is usually assumed in the literature.  We
comment below on the implications of this for moduli stabilisation.

Before looking at the moduli stabilisation and de Sitter vacua, 
we present a second representative model, which has a single hidden sector gauge group.

\subsubsection{Model II: single gaugino condensate}

The shift vector, Wilson lines, massless spectrum and all other details of this model  
are given in Appendix~\ref{app:Model2}. The unbroken gauge group after compactification 
is $\SU{3}\x\SU{2}\x\U1_Y \x\left[\SO{14}\x\U1^5\right]$.
The modular symmetry is broken by the Wilson lines to 
$SL(2,\mathbb{Z}) \times \Gamma_1(3) \times \Gamma_1(2) \times
\Gamma^1(2)$, where the transformations act on $T_1,T_2,T_3$ and $U$ respectively.  
 
The computation of the action follows exactly as for Model I, and
again we relegate the details to the appendix.

\paragraph{K\"ahler potential.}

The K\"ahler potential is given by~\eqref{Kahler1}, where the modular
weights can be found in Table~\ref{tab:4DSpectrumModel2}, and the
Green-Schwarz coefficients are
\begin{equation}
  \label{deltaGSsII}
  \delta_{GS}^1 = - 11 \,,\quad \delta_{GS}^2 =-\frac{11}{3}\,,\quad \delta_{GS}^3 =0\,.
\end{equation}
Again, we neglect contributions to $K$ from the singlet {\it vevs}, 
since they are small.

\paragraph{Gauge kinetic function.}

The beta-function coefficients that measure the field theoretic threshold corrections
are
\be
 b_{SO(14)} = -29, \qquad \, b^0_{SO(14)} = -36 \, .
\ee
Thus, the gauge kinetic function for the single hidden sector is given by 
\be\label{f14}
f_{14} = S - \frac{7}{8\pi^2} \log{\frac{M_d}{M_{P}}} +  \frac{7}{6\pi^2}
\, \log\eta\left(3\, T_2\right)^2 + \frac{5}{8\pi^2} \left(
 \log\eta\left(2\,T_3 \right)^2  + \log\eta\left(U/2\right)^2\right) \, .
\ee
Using the data in Appendix~\ref{app:Model2}, we can check that the 
constraint \eqref{matconst} is satisfied, the expressions are modular
covariant, and the universal $\delta_{GS}^j$ are obtained.

\paragraph{Superpotential.}

As in the previous model, we consider the contributions to the superpotential
from Yukawa couplings and gaugino condensation. The allowed Yukawa
couplings for this model are computed in Appendix~\ref{app:Model2}.  
As before, couplings among singlets are
all between $\vartheta^2\vartheta^5\vartheta^5$ sectors, and the
expressions for the couplings in terms of the  $T_i$ moduli can be
found in Appendix~\ref{app:Yukawas}.  Using the notation in the
appendix, we can have couplings of four types. 

We can compute the holomorphic superpotential and check that the
expressions are modular covariant.   
In what follows, we
consider only the  following terms~\footnote{These come from the
  prototype couplings  $n_7 n_{68} n_{69}$,  
$n_{17} n_{68} n_{69}$,   $n_{7} n_{61} n_{62}$,  $n_{17} n_{61}
  n_{62}$ (see Table~\ref{tab:3pointCouplingsModel2}),  by taking
  consistently the {\it vevs} $n_{61} = n_{62} = n_{68} = n_{69} = A_1$, $n_7 = A_2$, 
$\sqrt2 n_{17} = A_3$. We have also turned on other allowed couplings, but the results were not more interesting.} 
\be
  \label{WyukModel2}
  W_{\it yuk} = 2 \, N_{255}\, A_1^2 \,  \left(A_2 
  +A_3 \, e^{-2\pi \, T_1/3}\right) \,,
   \ee
where we have considered only the leading instanton contributions, 
and replaced the singlets by their {\it vevs}. Notice the important
fact that a constant piece has naturally arisen, coming from the
leading contribution to the Yukawa coupling 
between twisted fields localised at the same fixed point
(see Appendix
\ref{app:Yukawas255}).

Finally, the gaugino condensation contribution to the superpotential is simply, 
\be\label{Wmodel2}
W_{\it gc} = -\frac{c}{e}\,\frac{3}{8 \pi^2}\,\, e^{-2\pi^2 S/3} \, \left(\frac{M_d}{M_P}\right)^{7/12} 
 \, \eta(3\,T_2)^{-14/9} \, \left[ \eta(2\,T_3)\,\eta(U/2)\right]^{-5/6}
 \,.
\ee
Notice that in this case, the Dedekind eta functions do appear with
negative powers in the superpotential, 
as is usually assumed.

\subsection{Towards stabilisation of moduli and de Sitter vacua}

In this section, we study the problem of moduli stabilisation and de
Sitter vacua in the two explicit orbifold models discussed  in
the previous section.  

The scalar potential that governs
the dynamics of the moduli, after having integrated out the hidden matter,
is given to a good approximation (in Planck units) by
\be\label{FinalV}
V_F= e^K \, |W|^2 \Bigg[   \left| Y\,  \frac{W_S}{W} - 1 \right|^2 
  + \sum_j \frac{Y}{Y - \delta^j } \,
\left| \delta^j\, \,\frac{W_S}{W} +  (\phi_j + \bar {\phi_j}) \, \frac{W_{\phi_j}}{W}  - 1 
 \right|^2 - 3  \,\,  \Bigg]\,,
 \ee
where, as before,
$Y = S+ \bar{S} - \sum_j \delta^j\,  \log{(\phi_j + \bar {\phi_j})}$ with 
$\phi_j = T_1, T_2, T_3, U$,  $\delta^j = \frac{\delta_{GS}^j}{8\pi^2}$,   and now $W= W_{yuk} + W_{gc}$. 
We have to analyse this potential for the five complex fields $S, T_1,
T_2, T_3$ and $U$. Due to the complexity of the system, we can only
make limited progress analytically, in particular we can solve for the
axionic parts.  Thereafter, our search for vacua will be numerical.
First, however, we can make some general observations.

\subsubsection{de Sitter vacua}
Before considering  the moduli stabilisation mechanisms that may be at
work, let us begin by discussing the structure of the potential, and
in particular the possibility of obtaining de Sitter solutions.  
From (\ref{FinalV}) we see that the positivity of the potential
depends basically on how large the ratios $W_a/W$ can be, compared to
one.  If three of our five fields have $W_a/W$ sufficiently smaller or
larger than one, then the potential will certainly be positive. In
order for it to be negative, the ratio $W_a/W$  has to be of order
one for at least three of the fields. One can then analyse the
dependence of the ratio $W_a/W$ for each field.  
For this, let us recall the orders of magnitude that we are considering.  
{\it i)} The {\it vevs} of the matter fields must be around 
$A_\alpha \lesssim 10^{-1}$ in order for the calculation of the K\"ahler potential to be
valid.   {\it ii)} The decoupling scale must be lower than the Planck scale and thus
we have $M_d \lesssim  10^{-1} M_P$.  

For Model II, it is simple to estimate that the contribution from the
dilaton will generically give rise to a constant bigger than one, and
thus can cancel a big part of the negative piece in the
potential. Moreover, for $T_1$, the ratio $W_a/W$ can be seen to be
much less than one, thus contributing again to the positivity of
$V$. For the other fields, it depends on $\eta'$ and thus can in
principle be of order one and therefore give a null contribution to
the potential.   
The upshot is that for Model II we can generically expect $V>0$, and thus, de Sitter
vacua. An analogous analysis can be made for Model I, with similar
results.  
Note that the loop contribution from $\delta^j$ is always smaller than
the others, as it must be.  

We now ask whether or not we can expect to find vacua with all the
moduli stabilised.

\subsubsection{The modular form mechanism for $T_2, T_3, U$}

Let us first focus on the three moduli, $T_2, T_3, U$, which appear in
the superpotential due to stringy threshold corrections, {\it via} the
Dedekind eta functions defined in (\ref{eq:Dedekind}).  It was 
suggested in \cite{FILQ} that such a dependence may be enough to force
compactification and moduli stabilisation.  In particular, if the
eta functions appear in the superpotential with negative powers, then
the scalar potential diverges as the relevant moduli
go to zero and infinity (this can be seen by recalling the 
asymptotic behaviour of the eta function described in Section
\ref{S:modinv}).  Then a minimum in these directions is guaranteed.
Moreover, with the simplest factorisable superpotentials such as those
corresponding to a single gaugino condensate only, the fixed
points under the modular transformations are necessarily extrema of
the potential 
\cite{FILQ, BLST1}. 

Although the latter property is not observed for our models, which
have more complicated superpotentials, Model II does have the required
asymptotic behaviour to ensure the existence of minima in the
$T_2,T_3,U$ directions.
However, in Model I, which is also  a very common one  in the landscape, the Dedekind eta functions appear in the
superpotential with positive powers, so that the scalar potential goes
to zero as the relevant moduli go to zero and infinity.  In that case,
we cannot ensure the existence of minima.
Notice also that in realistic orbifolds like the ones we consider
here, Wilson lines break the modular symmetries in
each plane in different ways, and thus
generically, it is not correct to assume equality of all the K\"ahler
moduli 
(the dependence of the superpotential on each K\"ahler modulus assumes
a different functional form).  

The existence of multiple gaugino condensates and how the threshold corrections enter the superpotential
are somehow correlated with each other.  If
the $E_8 \times E_8$ gauge group is broken down to yield multiple
non-Abelian hidden gauge groups, the latter will tend to be small, and
thus one can expect to find large amounts of hidden matter.  
At the same
time, much hidden matter, whether massless or massive, drives the
auxiliary ${\mathcal N}=2$ beta-function coefficients towards positive values, which in turn means
that the eta functions appear with positive powers in $W_{gc}$.

The case of $T_2$ in Model I is slightly different since it also
always appear in the Yukawa couplings.  Thus a different mechanism may
help to stabilise this field.

\subsubsection[Racetrack and KKLT for $S, T_1, T_2$]{Racetrack and KKLT for $\bs{S, T_1, T_2}$}

 The dependence of the potential on $T_1$ can occur only via the
 non-perturbative Yukawa couplings in \Z6-II orbifolds.
 The form of the superpotential is then reminiscent of the KKLT one
 \cite{KKLT}, albeit 
 with coefficients that depend on the other moduli.  The same is true
 for $S$, in the case of a single gaugino condensate.  Otherwise, $S$
 may be stabilised with a racetrack induced by multiple 
 gaugino condensates (see \cite{CCM}).  Similarly, if
 $T_2$ appears in $W_{yuk}$, it may have a racetrack behaviour, with
 $W_{yuk}$ competing against the leading dependence in $W_{gc}$.

Since the standard racetrack and KKLT stabilisation mechanisms both
require some degree of tuning of the coefficients, and since in our
case those
coefficients are actually field-dependent, whether or not they are
successful in producing a minimum in all directions must be studied
explicitly.  We now turn to that analysis.

\subsubsection{Stabilisation of the axions}

We can make significant progress by first considering the imaginary
parts of our five complex fields, the axions.  These turn out to have periodic
potentials, with multiple minima and maxima, several of which can be
found analytically.  To begin with, recall that the axions contribute 
to the scalar potential only via the superpotential.

For the axions entering the potential {\it via} the Dedekind eta
functions, it is in general difficult to identify the positions of all
the extrema. For example, only for the simplest superpotentials do the
fixed points correspond to extrema.  However, it turns out that ${\rm
  Im} \,\phi_j = 0$ are typically minima, and so we may choose those
points.  This simplifies the problem for the remaining fields~\footnote{
These extrema are independent of the real components of $\phi_j$.
It is however clear that there are other critical points ${\rm Im} \,\phi_j$ 
which do depend on the specific values of ${\rm Re} \,\phi_j$.}. For
example, consider the superpotential for Model II, which can now be
written in the form
\be 
W = C_1 \, e^{-c_1 T_1} + C_2(T_2,T_3,U) \,e^{-c_2 S} + C_3 
\ee
with all the coefficients real.  This is
reminiscent of a KKLT superpotential \cite{KKLT} for the fields $T_1$
and $S$, although here the KKLT coefficients are no longer constants,
but depend on the other fields.  For the dependence on $T_2, T_3$ and
$U$, let us consider just the leading terms in the eta functions.
Then the superpotential is simply a sum of exponentials, and the
derivatives that appear in the scalar potential are easy to compute.

Similar to the basic KKLT model, the resulting scalar potential
depends on ${\rm Im} \,T_1=\sigma_1$ and ${\rm Im} \,S = \sigma_S$ as
terms proportional to $\cos c_1 \sigma_1$, $\cos c_2 \sigma_S$ and
$\cos (c_1 \sigma_1 - c_2 \sigma_S)$.  The magnitudes and signs of the
coefficients depend on $C_1, C_2, C_3$ as well as the real parts of
the five complex fields.  In any case, it is clear that $\sigma_1 =
n_1 \pi/c_1$ and $\sigma_S = n_2 \pi/c_2$, with $n_1, n_2$ integers,
are critical points, and so they represent interesting candidates for
the minima.  Notice that analogous arguments can be made for models
that fall into the racetrack class.

With extrema for the five imaginary parts in hand, we now turn to a
numerical analysis for the real parts.  We have also scanned
  numerically for all ten real and imaginary parts together. 

\subsubsection{Numerical results  and explicit dS vacua}

We have tried several methods to search numerically for minima in the
scalar potential~\footnote{We especially thank Yashar Akrami, David
  Grellscheid, Stuart Raby and Timm Wrase for
  detailed discussions on this point.}. The most successful one we found was to search
for solutions to 
the critical point equations~\footnote{We have performed numerous
  checks to our solutions {\it e.g.} minimizing numerically the one
  dimensional problems, examining the 2D and 3D plots, and increasing the numerical precision to up
  to 1000 figures.  Note that not all candidate critical points 
that one finds survive these checks.  For example, it can happen that the 
initial numerical computation finds minima in two directions that are 
slightly shifted from each other, and does not resolve this difference.} 
$\partial V/\partial x^A = 0$ for $x^A$ 
the real (and imaginary) parts of $S,T_i,U$. 

By varying the input parameters $c,A_\alpha$ and $M_d$ we have identified several critical points consistent
with our approximations~\footnote{In addition, we have taken the first 20
terms in the expansion of the eta function. We have verified that higher 
order terms do not alter our results.} in both models. In Tables~\ref{unstabledSI} 
and~\ref{unstabledSII} 
we show some representative vacua of Model I and Model II,
respectively.  Mathematica codes with all the details of these vacua
can be found at our webpage \cite{web}. 

\begin{table}[t]
\centering
{\small
\begin{tabular}{|c|c|c|}
\hline
\multicolumn{3}{|c|}{Parameters}\\
$c=1/10$ & $20A_1=100A_2=A_3=1/10$ & $M_d=1/65$ \\ 
\hline
\hline
\multicolumn{3}{|c|}{Moduli {\it vevs}}\\
$\phi_i$ & ${\rm Re}\langle\phi_i\rangle$ & ${\rm Im}\langle\phi_i\rangle$ \\
\hline
$T_1$ & $3.51166$ &  $3/2$ \\ 
\hline
$T_2$ & $0.24201$ &  $-1/3$ \\
\hline
$T_3$ & $7.05981$ &  $35/8$ \\
\hline
$U$ & $112.95695$ &  $-506$ \\
\hline
$S$ & $0.09385$ &  $1079/144\pi$ \\
\hline
\hline
\multicolumn{3}{|c|}{Cosmological constant $\Lambda = 6.45\x10^{-19}$}\\
\hline
\multicolumn{3}{|c|}{Effective dilaton $Y = 0.32262$}\\
\hline
\hline
\multicolumn{3}{|c|}{Mass eigenstates}\\
$\rho_i\sim \phi_i$ & $m^2_{{\rm Re}\rho_i}$ & $m^2_{{\rm Im}\rho_i}$ \\
\hline
$\rho_1$ & $1.87\x10^{-18}$ & $3.57\x10^{-18}$ \\
\hline
$\rho_2$ & $-4.76\x10^{-17}$ & $1.2\x10^{-17}$ \\
\hline
$\rho_3$ & $3.15\x10^{-20}$ & $3.29\x10^{-33}$ \\
\hline
$\rho_U$ & $9.84\x10^{-23}$ & $1.51\x10^{-94}$ \\
\hline
$\rho_S$ & $1.14\x10^{-16}$ & $2.16\x10^{-16}$ \\
\hline
\multicolumn{3}{|c|}{Tachyon: ${\rm Re}\rho_2\sim0.9{\rm Re}T_2+0.4{\rm Re}S$}\\
\hline 
\end{tabular}
}
\caption[]{An unstable de Sitter solution of Model I. The dominant contribution to the 
mass eigenstate $\rho_i$ arises from the modulus $\phi_i$. In particular, 
the tachyon is dominated by the ${\rm Re}\,T_2$ direction, 
as can also be appreciated in Figure~\ref{fig:ModelIplots}. 
We give the solution to 5 significant figures, but we 
have computed it to a precision of 1000. The value of $\Lambda$ is given in Planck units.}
\label{unstabledSI}
\end{table}

\FIGURE[t!]{\epsfig{file=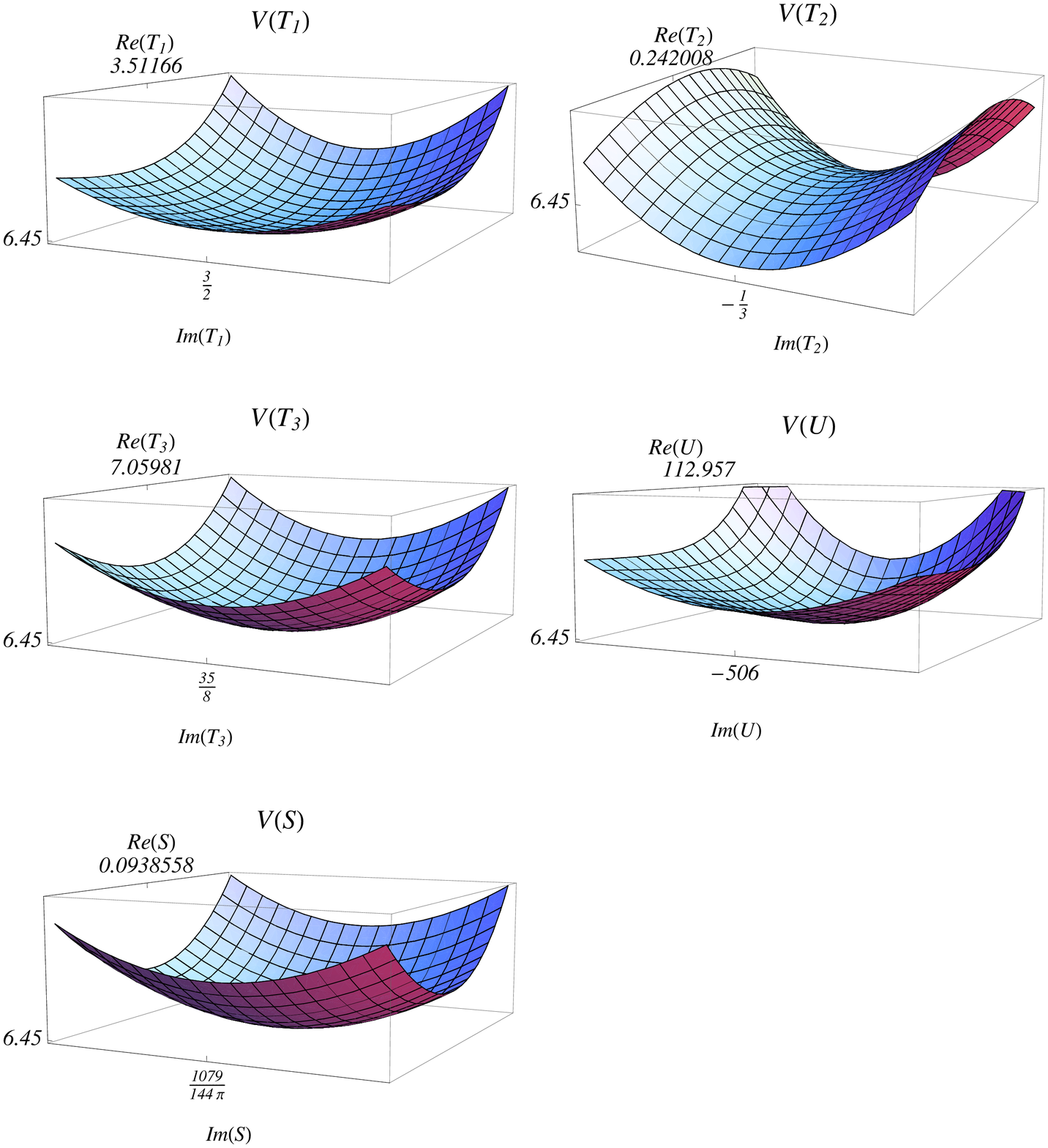,width=\linewidth}
 \vspace{-2cm}
 \caption[]{Scalar potental as a function of the bulk moduli of Model I for the vacuum
  presented in Table~\ref{unstabledSI}. The scalar potential is rescaled by a factor $10^{19}$.}
 \label{fig:ModelIplots}}

\begin{table}[t]
\centering
{\small
\begin{tabular}{|c|c|c||c|c|c|}
\hline
\multicolumn{6}{|c|}{Parameters}\\
$c=1/2$ & $A_1,A_2,A_3=1/2$ & $M_d=1/600$ & $c=1/200$ & $A_1,A_2,A_3=1/2$ & $M_d=10^{-3}$\\ 
\hline
\hline
\multicolumn{6}{|c|}{Moduli {\it vevs}}\\
$\phi_i$ & ${\rm Re}\langle\phi_i\rangle$ & ${\rm Im}\langle\phi_i\rangle$  & $\phi_i$ & ${\rm Re}\langle\phi_i\rangle$ & ${\rm Im}\langle\phi_i\rangle$\\
\hline
$T_1$ & $0.60565$ &  $3/2$  & $T_1$ & $0.60213$ &  $3/2$\\ 
\hline
$T_2$ & $2.05917$ &  $19/6$ & $T_2$ & $3.99948$ &  $19/6$\\
\hline
$T_3$ & $6.48601$ &  $37/4$ & $T_3$ & $11.95957$ &  $37/4$\\
\hline
$U$ & $25.94402$ &  $-3$ & $U$ & $47.83828$ &  $-3$\\
\hline
$S$ & $0.53034$ &  $809/144\pi$ & $S$ & $0.97758$ &  $809/144\pi$\\
\hline
\hline
\multicolumn{3}{|c||}{Cosmological constant $\Lambda = 7.41\x10^{-6}$} & \multicolumn{3}{c|}{$\Lambda = 5.12\x10^{-7}$}\\
\hline
\multicolumn{3}{|c||}{Effective dilaton $Y = 1.15312$} & \multicolumn{3}{c|}{$Y = 2.07762$}\\
\hline
\hline
\multicolumn{6}{|c|}{Mass eigenstates}\\
$\rho_i\sim \phi_i$ & $m^2_{{\rm Re}\rho_i}$ & $m^2_{{\rm Im}\rho_i}$ & $\rho_i\sim \phi_i$ & $m^2_{{\rm Re}\rho_i}$ & $m^2_{{\rm Im}\rho_i}$\\
\hline
$\rho_1$ & $2.88\x10^{-5}$ & $2.28\x10^{-4}$ & $\rho_1$ & $2.57\x10^{-6}$ & $2.02\x10^{-5}$\\
\hline
$\rho_2$ & $-1.67\x10^{-5}$ & $8.34\x10^{-19}$ & $\rho_2$ & $-1.92\x10^{-7}$ & $8.61\x10^{-36}$\\
\hline
$\rho_3$ & $5.57\x10^{-7}$ & $1.21\x10^{-37}$ & $\rho_3$ & $1.2\x10^{-8}$ & $1.23\x10^{-68}$\\
\hline
$\rho_U$ & $2.84\x10^{-8}$ & $7.57\x10^{-39}$ & $\rho_U$ & $5.97\x10^{-10}$ & $7.72\x10^{-70}$\\
\hline
$\rho_S$ & $4.27\x10^{-4}$ & $7.87\x10^{-4}$ & $\rho_S$ & $5.28\x10^{-5}$ & $6.72\x10^{-5}$\\
\hline
\multicolumn{3}{|c||}{Tachyon: ${\rm Re}\rho_2\sim0.9{\rm Re}T_2+0.3{\rm Re}T_3+0.4{\rm Re}S$} & \multicolumn{3}{c|}{${\rm Re}\rho_2\sim0.9{\rm Re}T_2+0.4{\rm Re}T_3+0.2{\rm Re}S$}\\
\hline 
\end{tabular}
}
\caption[]{Unstable de Sitter solutions to Model II. As in Model I, the dominant contribution to the tachyon
arises from ${\rm Re}T_2$. The potential $V({\rm Re}T_2)$ 
differs from
the one for Model I because the tachyon has some components in some other real directions. 
A change in the input parameters provides a larger value for $Y$.
We give the solution to 5 significant figures, but we have computed it to a precision of 1000.}
\label{unstabledSII}
\end{table}

The vacua we find are all de Sitter, however they all turn out to have at least one unstable direction in the moduli space. This is evident from the plot of the scalar potential
as a function of ${\rm Re}\,T_2$ in Model I, and/or from the mass eigenvalues,
displayed in Tables~\ref{unstabledSI} and \ref{unstabledSII}. 
One might be tempted to
think that  a shift to the left in $V({\rm Re}\,T_2)$ of Model I leads to the 
true minimum. However, this is not sufficient
since any change in ${\rm Re}\,T_2$ affects
the whole scalar potential, resulting in a vacuum with (more) instabilities in
other directions.
In general, we have found it very difficult to avoid instabilities in the direction ${\rm Re}\,(T_2,\,T_3,\,U)$.
It would be important to understand the reason for these
instabilities, and whether or not they are inevitable.

An interesting feature of both models is that there are some almost
flat directions, as observed from the hierarchically small masses.
The corresponding eigenvectors are linear combinations of ${\rm Im}\, S, {\rm Im} \,T_3$ and
${\rm Im} \,U$ 
(and also ${\rm Im}\, T_2$ for Model II).  The reason for these almost flat directions is as
follows.  First, notice that in Model I, the contribution to $W_{gc}$
from the \SU3 gaugino condensate is typically highly suppressed
with respect to the \SO8 condensate due to the stronger
exponential suppression in $S$, and the stronger suppression in
$M_d/M_P$.  Thus, effectively, we have only a single condensate in
both models.  Moreover, the higher order contributions to the eta
functions in both Model I and Model II, at their solutions, are
strongly suppressed for $T_3$ and $U$ (the same can be said for
$T_2$, when ${\rm Re}\,T_2$ becomes larger than one as in Model II).  
Therefore, to a good approximation, ${\rm Im} \,S, {\rm Im} \,
T_3$ and ${\rm Im} \,U$ (and ${{\rm Im} \,T_2}$ for Model II) appear in the superpotential, and thus in the
scalar potential, only via a single linear combination, which, {\it
  e.g.} for Model I, can be written as 
\be 
X~=~ -\frac{4\pi^2}{3} {\rm Im}\,S - \frac{\pi}{9} {\rm Im} \,T_3 -
\frac{\pi}{144} {\rm Im}\, U \,.  
\ee
As a consequence, to a good approximation, only this linear
combination of the three (or four) fields is lifted by the non-perturbative
dynamics, leaving two (or three) linearly independent combinations as flat directions, and two (or three) corresponding shift symmetries.  Of course, the
higher order corrections do lift the latter directions, rendering them
only almost flat.  We expect this feature to be also present in
any metastable minima that may exist, being quite a generic
consequence of the moduli stabilisation via gaugino condensation with
threshold corrections.  Such almost flat directions could potentially
be interesting, since extremely light scalar fields are often called
upon in cosmological models, for inflation, quintessence and the like.

Finally, we comment on the {\it vevs} of the moduli that we find.  Of
course, due to the instabilities, our vacua cannot be considered as
candidates for our universe, and so we have not attempted to obtain
realistic values for the moduli at this stage.  However, consistency of our analysis
does impose some constraints.  In particular, the overall volume must
be large enough for the supergravity approximation to hold, and the
dilaton large enough for the string loop expansion to be a good one.
The latter is difficult to achieve, as has long been known in
heterotic orbifolds.
Meanwhile, it is interesting that anistropic compactifications 
emerge quite naturally in our setup, since 
these have been proposed to
explain the mild hierarchy between the GUT and string scales.

\section{Discussion}\label{Concl}

We have revisited the problem of moduli stabilisation in heterotic
orbifold compactifications, in the light of the recent discovery of
fertile patches in the heterotic landscape where the MSSM spectrum can
be found.  After reviewing the derivation of the low energy effective
action describing general orbifold compactifications, we computed it
explicitly for two concrete MSSM models.  Our resulting actions survived several
non-trivial checks, such as modular invariance, including a
new stringy constraint on the matter content in the presence of
gaugino condensation (\ref{matconst}).  Finally, we  studied
the dynamics for all the bulk moduli fields, making use of numerical
techniques.  We found several de Sitter vacua, but unfortunately they
all suffer from instabilities.  Curiously, we did not uncover any
anti-de Sitter vacua that were consistent with our assumptions. 
Since the scalar potential grows rapidly, we expect them 
to appear at very small moduli {\it vevs}, which are not consistent with
our approximations.
On the other hand,  the values of the cosmological constant at the
vacua tend to be very large, although we should recall that this
is counting only the classical value for the vacuum energy.

 Given that we have found it difficult to locate 
metastable vacua, it is interesting to compare our results with previous
toy models for heterotic orbifold compactifications.  The most recent
of these is \cite{DRW}, which treated two bulk moduli, $S$ and $T$,
together with a complex matter field.  Although they included a D-term
contribution, it turns out to be vanishingly small, so that their
setup is very similar to ours.  Indeed, it is easy to find metastable
minima to their F-term potential.  Thus it seems that the challenge we
meet comes from considering all five bulk moduli fields that are present and the comparatively small number of free parameters.

Our analysis leads us to question if new ingredients, beyond those
already considered in the literature, may well be necessary in order
to stabilise all the moduli.  Along those lines, recall that we have
neglected some moduli-dependent contributions to the effective
Lagrangian, which are known to be present, but have been little
studied.  Firstly, there is a universal contribution to the threshold
corrections \cite{Omega}, and secondly an overall factor in the Yukawa
couplings, which can depend on bulk moduli of the untwisted planes
\cite{Erler:1992gt}.  It would also be very useful to have in hand
direct and explicit computations of the threshold corrections in the
presence of discrete Wilson lines \cite{ThresWL}. 
Other effects that depend on the moduli and merit further study are those arising from
the Casimir Energy and Coleman-Weinberg potential, which become
relevant after supersymmetry breaking (see {\it e.g.}~\cite{Casimir}).

Otherwise, the main unknown in our analysis has been the dynamics that
decouples exotics and hidden matter, which we parametrised with
some (non-Abelian) singlet {\it vevs} and the decoupling mass scale.  The
introduction of this high scale, which separates the dynamics of the
matter and moduli, is important for several reasons.
Phenomenologically, we know that the exotics must be
decoupled at least at the GUT scale to maintain gauge coupling unification, and yet the gauge hierarchy
problem suggests supersymmetry breaking should occur down at the TeV scale.
Practically, it renders the system tractable; indeed already with 10
real bulk moduli the search for minima has proven to be a significant
challenge.  Occasionally, it is also important because hidden matter,
if not massive, tends to destroy asymptotic freedom in the hidden
gauge groups, preventing the gauginos from condensing there.

We remark that in considering only the bulk moduli stabilisation, we
have assumed that all the other flat directions, corresponding to
twisted fields, may be lifted in the decoupling process, and this
certainly deserves more attention. 
One can argue that, since the orbifold MSSM candidates 
arise from ${\mathcal N}=(0,2)$ models, the massless
twisted fields might not generically correspond to blow-up modes
\cite{DFMS}; thus it should be possible to stabilise them and remain
within the orbifold description where we have much more control.  A
detailed treatment of the decoupling dynamics would require, among
other things, explicit computations of the higher order couplings
between non-Abelian singlet fields, and non-Abelian singlets and charged
matter.  For progress in this direction see~\cite{choiHO}.  Notice
that if a decoupling scale is in play, D-term potentials are no longer
relevant for the bulk moduli dynamics, since all the \U1 gauge groups
are broken when the non-Abelian singlets acquire {\it vevs} (all the
non-Abelian singlets are generically charged under all the hidden \U1's).

Although unstable, the dS vacua we have found do have some interesting properties.
In particular, the non-perturbative dynamics ({\it i.e.} gaugino
condensation with threshold corrections) tends to give rise to some
almost flat directions, and consequently hierarchically light fields.
We expect that this feature can also hold for any metastable dS vacua
that might exist. Light fields could be useful for cosmology, and
it would then be important to understand how quantum corrections
contribute to the light masses after supersymmetry breaking.

So far we must say that our results on the (non-)existence of
metastable de Sitter in MSSM heterotic orbifolds are inconclusive.  We
are now developing more sophisticated numerical codes to scan our
scalar potentials for metastable minima, and are widening our search
in the minilandscape of MSSM models \cite{yashar}.  Then we must also
ask if a model can be found with realistic values for the moduli {\it
  vevs}, including Re $S \sim 2$, an appropriate gravitino mass and
cosmological constant.  
If and when such a vacuum is found,
it would have a host of applications.  We could begin to study in
detail the phenomenological and cosmological aspects of the
compactification, starting with supersymmetry breaking and the low
energy sparticle spectrum, models of inflation, resolutions to the
cosmological moduli problem and so on.

\section*{Acknowledgements}
We are grateful to L.~Velasco-Sevilla for collaborations at various stages of this project. 
We would also like to thank Y.~Akrami, A.~Ashoorioon, K.~Bobkov, N.~Cabo-Bizet, K.~S.~Choi, U.~Danielsson,
B.~Dundee, A.~Ernvall-Hyt\"onen, S.~F\"orste, R.~Gregory, D.~Grellscheid, T.~Kobayashi, 
O.~Lebedev, J.~Louis, C.~L\"udeling, A.~Micu, H.~Monien, H.P.~Nilles, N.~Pagani, S.~Raby, F.~Quevedo,
G.~Tasinato, T.~Van Riet, P.~Vaudrevange and T.~Wrase for interesting and helpful
discussions. S.~R-S. is thankful to IPMU 
and KITP, where part of this work was done, for hospitality and support.
S.~R-S. is particularly grateful to T.~Yanagida and to the organisers 
of the program ``Strings at the LHC and in the early universe'' 
for their invitation and discussions.   S.L.P is supported by the
G\"{o}ran Gustafsson Foundation.
I.~Z.~was supported by the DFG cluster of excellence
Origin and Structure of the Universe, the SFB-Tansregio TR33
``The Dark Universe" (Deutsche Forschungsgemeinschaft) and
the European Union 7th network program ``Unification in the
LHC era" (PITN-GA-2009-237920).

\appendix

\section{Details of Model I}
\label{sec:Model1}

In this appendix we describe in detail our first explicit MSSM orbifold compactification.  We specify the orbifold, and present its complete 4D massless spectrum, along with the trilinear couplings allowed by the standard selection rules.  Then we describe the modular symmetries and anomalies, and compute the coefficients that measure the stringy threshold corrections to the gauge kinetic functions.

\subsection{Model definition}
\label{app:Model1def}

The model is defined by the following shift vector and Wilson lines
\begin{eqnarray}
  \label{eq:ModelVandW}
  V^{\text{SO(10),1}} & = &
    \left(\tfrac13,\,-\tfrac12,\,-\tfrac12,\,0,\,0,\,0,\,0,\,0\,\right) \left( \tfrac12,\,-\tfrac16,\,-\tfrac12,\,-\tfrac12,\,-\tfrac12,\,-\tfrac12,\,-\tfrac12,\,\tfrac12\,\right)\,,\\
 {\cal A}_5 &= &
    \left(0,\,\tfrac{1}{2},\,\tfrac{1}{2},\,\tfrac{1}{2},\,-\tfrac{1}{2},\,0,\,0,\,0\,\right) \left( 1,\,0,\,0,\,0,\,2,\,-2,\,0,\,0\,\right)\,,\nonumber\\
 {\cal A}_3 &= &
   \left(0,\,-1,\,-\tfrac13,\,-\tfrac13,\,-\tfrac13,\,-\tfrac13,\,-\tfrac13,\,\tfrac23\,\right) \left( \tfrac{10}{3},\,-2,\,-4,\,-\tfrac73,\,-\tfrac73,\,-4,\,-3,\,3\,\right)\,.\nonumber
\end{eqnarray}
The Wilson lines ${\cal A}_3$ (order 3) and ${\cal A}_5$ (order 2) lift the
degeneracy of the fixed points in the second and third torus,
respectively. A possible second order 2 Wilson line in the third torus
is set to zero.  The 4D gauge group after compactification is
\begin{equation}
  \label{eq:4dgg}
 \SU{3}_C\x\SU{2}_L\x\U1_Y\x\left[\SO{8}\x\SU{3}\x\U1^6\right]\,.
\end{equation}
The \U1 generators are chosen to be
\begin{subequations}
\begin{eqnarray}
\label{eq:U1generators}
 t_1 & = & t_{\rm X}=\left(\tfrac53,\,1,\,-\tfrac53,\,\tfrac13,\,\tfrac13,\,\tfrac13,\,\tfrac13,\,\tfrac13\,\right) \left(-\tfrac13,\,-\tfrac73,\,0,\,\tfrac13,\,\tfrac13,\,0,\,0,\,0\,\right)\,,\\
 t_2 & = & t_{\rm Y}=\left(0,\,0,\,0,\,\tfrac{1}{2},\,\tfrac{1}{2},\,-\tfrac{1}{3},\,-\tfrac{1}{3},\,-\tfrac{1}{3}\,\right) \left( 0,\,0,\,0,\,0,\,0,\,0,\,0,\,0\,\right)\,,\\
 t_3 & = &\left(-45,\,321,\,45,\,-9,\,-9,\,-9,\,-9,\,-9\,\right) \left( 9,\,63,\,0,\,-9,\,-9,\,0,\,0,\,0\,\right)\,,\\
 t_4 & = &\left(246,\,0,\,75,\,-15,\,-15,\,-15,\,-15,\,-15\,\right) \left( 15,\,105,\,0,\,-15,\,-15,\,0,\,0,\,0\,\right)\,,\\
 t_5 & = &\left(0,\,0,\,171,\,15,\,15,\,15,\,15,\,15\,\right) \left( -15,\,-105,\,0,\,15,\,15,\,0,\,0,\,0\,\right)\,,\\
 t_6 & = &\left(0,\,0,\,0,\,3,\,3,\,3,\,3,\,3\,\right) \left( 54,\,-21,\,0,\,-54,\,-54,\,0,\,0,\,0\,\right)\,,\\
 t_7 & = &\left(0,\,0,\,0,\,21,\,21,\,21,\,21,\,21\,\right) \left( 0,\,15,\,0,\,0,\,0,\,0,\,0,\,0\,\right)\,,
\end{eqnarray}
\end{subequations}
where $t_Y$ and $t_X$ denote respectively the hypercharge and the anomalous \U1 generator.

\subsection{4D spectrum}
\label{app:4Dspectrum}

The spectrum in the
untwisted sector includes: the graviton, the axio-dilaton, $S$, the gauge bosons of the gauge group~\eqref{eq:4dgg}, 
the K\"ahler moduli $T_i$, $i=1,2,3$, the complex structure modulus
$U_3$ ({\it cf.}
Section~\ref{sec:ConcreteExample}), and their corresponding supersymmetric partners. In addition, the
spectrum contains the matter fields with non-zero quantum numbers given
in Table~\ref{tab:4DSpectrumModel}.
They amount to three generations of quarks and leptons plus vector-like exotics with respect to
$\mathcal{G}_\mathrm{SM}$
and hidden matter.

{\footnotesize
\setlength{\LTcapwidth}{0.95\textwidth}
\begin{longtable}{|@{}c@{}|@{}l@{}|@{}rrrrrrr|@{}rrr|@{}r|l|}
\caption{The spectrum of the \Z6--II Model I.  We denote the fixed points, $f$,  in the $\vartheta^k$
  (un)twisted sector, with their (real) lattice coordinates.  A pair of coordinates $(-,-)$ indicates that the corresponding torus is left fixed
  in the sector.  We give the representations under $SU(3)_C \times SU(2)_L \times SO(8) \times SU(3)$ respectively, and the  \U1 charges $q_i$ with respect to the
  basis of generators given above.   $n^i$ and $\ell^i$ are the K\"ahler and complex structure modular weights in the $i$th torus.} 
\label{tab:4DSpectrumModel}
\\
\hline
\ \ \ $f$ & \ \ \ Repr. & $q_\mathrm{X}$ & $q_{Y}$ & $q_{3}$ & $q_{4}$ & $q_{5}$ & $q_{6}$ & $q_{7}$ & $n^1$ & $n^2$ & $n^3$ & $\ell^3$ & \!\!Label\!\! \\
\hline\hline
\endfirsthead
\hline
\ \ \ $f$ & \ \ \ Repr. & $q_\mathrm{X}$ & $q_{Y}$ & $q_{3}$ & $q_{4}$ & $q_{5}$ & $q_{6}$ & $q_{7}$ & $n^1$ & $n^2$ & $n^3$ & $\ell^3$ & \!\!Label\!\! \\
\hline\hline
\endhead
\hline
\multicolumn{14}{|r|}{{\it continued ...}$\phantom{_{6}^6}$}\\
\hline
\endfoot
\hline
\endlastfoot
\multicolumn{14}{|l|}{$\vartheta^0$ sector}\\
\hline
 & $(\boldsymbol{1}, \boldsymbol{1}, \boldsymbol{1}, \boldsymbol{1})$ & $-\frac{4}{3 }$ & $0$ & $-138$ & $-123$ & $123$ & $\frac{15}{2 }$ & $\frac{105}{2 }$ &  $-1$ & $0$ & $0$ &  $0$ & $n_{1}$\\
 & $(\boldsymbol{1}, \boldsymbol{1}, \boldsymbol{1}, \boldsymbol{1})$ & $\frac{2}{3 }$ & $-1$ & $156$ & $-168$ & $-78$ & $\frac{3}{2 }$ & $\frac{21}{2 }$ &  $-1$ & $0$ & $0$ &  $0$ & $\overline{e}_{1}$\\
 & $(\boldsymbol{\overline{3}}, \boldsymbol{1}, \boldsymbol{1}, \boldsymbol{1})$ & $\frac{2}{3 }$ & $\frac{2}{3 }$ & $156$ & $-168$ & $-78$ & $\frac{3}{2 }$ & $\frac{21}{2 }$ &  $-1$ & $0$ & $0$ &  $0$ & $\bar{u}_{1}$\\
\  & $(\boldsymbol{1}, \boldsymbol{1}, \boldsymbol{8}, \boldsymbol{1})$ & $\frac{7}{3 }$ & $0$ & $-63$ & $-105$ & $105$ & $21$ & $-15$ &  $0$ & $-1$ & $0$ &  $0$ & $\widetilde{N}_{2}$\\
 & $(\boldsymbol{3}, \boldsymbol{2}, \boldsymbol{1}, \boldsymbol{1})$ & $\frac{4}{3 }$ & $-\frac{1}{6 }$ & $-210$ & $78$ & $-78$ & $\frac{3}{2 }$ & $\frac{21}{2 }$ &  $0$ & $-1$ & $0$ &  $0$ & $q_{1}$\\
  & $(\boldsymbol{1}, \boldsymbol{2}, \boldsymbol{1}, \boldsymbol{1})$ & $2$ & $\frac{1}{2 }$ & $-54$ & $-90$ & $-156$ & $3$ & $21$ &  $0$ & $0$ & $-1$ &  $-1$ & $l_{1}$\\
 & $(\boldsymbol{1}, \boldsymbol{2}, \boldsymbol{1}, \boldsymbol{1})$ & $-2$ & $-\frac{1}{2 }$ & $54$ & $90$ & $156$ & $-3$ & $-21$ &  $0$ & $0$ & $-1$ &  $-1$ & $\overline{l}_{1}$\\
\end{longtable}
}

\subsection{Trilinear couplings}
\label{app:3pointCouplings}

The string selection rules allow us to identify all non-vanishing couplings present
in the effective theory emerging from an orbifold.  Here, in
Table~\ref{tab:3pointCouplings}, we give all trilinear couplings
allowed by the standard selection rules; gauge invariance, ${\cal R}$-charge conservation and the space group selection rule.

\begin{table}[!h!]
\begin{center}
{\footnotesize
\begin{tabular}{|c|cc|cc|c|c|}
\hline
$\vartheta^0\,\vartheta^0\,\vartheta^0$ & \multicolumn{2}{c|}{$\vartheta^3\,\vartheta^4\,\vartheta^5$} 
        & \multicolumn{2}{c|}{$\vartheta^2\,\vartheta^5\,\vartheta^5$} 
        & $\vartheta^2\,\vartheta^4\,\vartheta^0$ & $\vartheta^3\,\vartheta^3\vartheta^0$\\ 
\hline\hline
$\overline{u}_{1}\ \overline{l}_{1}\ q_{1}$ 
& $\overline{\delta}_{1}\ n_{30}\ \delta_{7}$ & $m_{1}\ n_{34}\ m_{15}$
& $\overline{d}_{1}\ \overline{d}_{6}\ \overline{u}_{2}$ & $n_{23}\ n_{59}\ n_{66}$
& $\overline{l}_{1}\ l_{2}\ n_{30}$ & $q_{1}\ m_{1}\ \overline{\delta}_{1}$ \\ 
& $\overline{\delta}_{3}\ n_{30}\ \delta_{7}$ & $m_{3}\ n_{34}\ m_{15}$
& $\overline{d}_{2}\ \overline{d}_{6}\ \overline{u}_{2}$ & $n_{9}\ s^+_{1}\ s^-_{3}$ 
& $\overline{l}_{1}\ n_{8}\ l_{5}$ 
& $q_{1}\ m_{2}\ \overline{\delta}_{2}$ \\ 
 & $\overline{\delta}_{1}\ n_{42}\ \delta_{7}$ & $m_{1}\ n_{47}\ m_{15}$
& $d_{1}\ \overline{e}_{2}\ \overline{u}_{2}$ & $n_{23}\ s^+_{1}\ s^-_{3}$
& $l_{1}\ n_{9}\ \overline{l}_{3}$ 
& $q_{1}\ m_{3}\ \overline{\delta}_{3}$ \\ 
 & $\overline{\delta}_{3}\ n_{42}\ \delta_{7}$ & $m_{3}\ n_{47}\ m_{15}$
& $d_{2}\ \overline{e}_{2}\ \overline{u}_{2}$ & $n_{9}\ s^+_{2}\ s^-_{5}$
& $\overline{l}_{1}\ l_{3}\ n_{42}$ 
& $q_{1}\ \overline{\delta}_{4}\ m_{6}$ \\ 
 & $m_{1}\ l_{5}\ s^-_{4}$ & $m_{1}\ \overline{l}_{3}\ s^+_{6}$
& $\overline{d}_{1}\ l_{7}\ q_{2}$ & $n_{23}\ s^+_{2}\ s^-_{5}$
& $l_{1}\ n_{20}\ \overline{l}_{4}$ 
& $q_{1}\ m_{7}\ \overline{\delta}_{5}$ \\ 
 & $m_{3}\ l_{5}\ s^-_{4}$ & $m_{3}\ \overline{l}_{3}\ s^+_{6}$
& $\overline{d}_{2}\ l_{7}\ q_{2}$ & $n_{9}\ \widetilde{N}_{67}\ \widetilde{N}_{67}$
& $\overline{l}_{1}\ n_{17}\ l_{6}$ 
& $q_{1}\ \overline{\delta}_{6}\ m_{10}$ \\ 
 & $m_{1}\ l_{6}\ s^-_{4}$ & $m_{1}\ \overline{l}_{5}\ s^+_{6}$
& $d_{1}\ q_{2}\ q_{2}$ & $n_{23}\ \widetilde{N}_{67}\ \widetilde{N}_{67}$
& $l_{1}\ \overline{l}_{2}\ n_{45}$ &  \\
\hline
\end{tabular}
\caption{All couplings at trilinear level allowed by the standard selection rules. We display all
116 couplings, classified according to the twisted sector in which the
involved fields appear.}
\label{tab:3pointCouplings}
}
\end{center}
\end{table}

\subsection{Modular symmetries}
\label{app:modsym}
The target space modular symmetry for the first plane is the
full group $SL(2,\mathbb{Z})$.  The Wilson lines in the second and
third planes
break the symmetry in those directions down to congruence subgroups, which
can be computed by solving the constraints Eqs. (48-55) found in the
first reference of
\cite{modWLBL}.  In the second plane the result is $\Gamma_1(3)$.  In the third plane, we have $\Gamma_0(4)$ acting on the K\"ahler
modulus, and $\Gamma^0(4)$ acting on the complex structure modulus.
Here we collect some of the properties of these groups.

The fundamental domain for $SL(2,\mathbb{Z})$ can be found in any text
book on modular forms.  It has a single cusp at infinity.  Representing the
modular transformations with the matrices
\be
\left( \begin{array}{cc} a & b \\ c & d \end{array}\right) \, ,
\ee
they are generated by
\be
{\mathcal S}: \quad  \left( \begin{array}{cc} 0 & -1 \\ 1 & 0 \end{array}\right), \qquad
{\mathcal T}:  \quad  \left( \begin{array}{cc} 1 & 1 \\ 0 & 1 \end{array}\right).
\ee
$SL(2,\mathbb{Z})$ has two self-dual points, also known as elliptic
points.  They can be computed by solving $\phi_* = \frac{a\phi_*
- ib}{ic\phi_*+d}$ for some $a,b,c,d$ and $\phi_*$.  It is convenient to
rotate $\phi$ so that its fundamental domain lies in the upper-half
complex plane, taking $\phi \rightarrow -i\phi$, in which case the
elliptic point is given by:
\be
\phi_* = \frac{a-d \pm \sqrt{(a+d)^2-4}}{2c} \, ,
\ee
after using $ad-bc=1$.  Here, $c\neq0$ and in order to lie within the
upper half plane, we must have $|a+d|<2$.  The solutions are at the
boundary of the fundamental domain, at (returning to our original
convention) $\phi_* = 1, \frac{\sqrt{3}}{2}-\frac{i}{2}$, which are fixed under $\mathcal{S},\mathcal{TS}$
respectively.  

Formally, the lowest weight modular form for $SL(2,\mathbb{Z})$ has
weight ${\rm k}=12$,
and is called the discriminant function, $\Delta$.  It is related to the Dedekind
eta function by $\Delta = \eta(\phi)^{24}$.  There are also modular forms with 
so-called character, which transform up to a phase factor, and the
Dedekind eta function with weight ${\rm k}=\frac12$ is one of these. 

Several results for the congruence subgroups can be found in the
textbooks.  Since these groups are contained
within $SL(2,\mathbb{Z})$, the above modular forms are also
modular forms of $\Gamma_0(N), \Gamma^0(N)$ and $\Gamma_1(N)$.
However, we also have additional ones, including $\eta(N\phi)$ for
$\Gamma_0(N),\Gamma_1(N)$.  Naturally, for a given $N$ the largest
space of modular 
forms will be for $\Gamma_1(N)$.  Generators for the groups can most easily
be computed by using the computer program SAGE, although the resulting
set is not small, nor even necessarily minimal.  For $\Gamma_1(3)$
we obtain a set of 12, of which we write down some examples:
\bea
\left( \begin{array}{cc} 1 & 1 \\ 0 & 1 \end{array}\right), \quad
\left( \begin{array}{cc} 1 & 0 \\  3 & 1 \end{array}\right), \quad 
\left( \begin{array}{cc} 1 & -1 \\  3 & -2 \end{array}\right). \quad 
\eea

It turns out that most congruence
subgroups do not have elliptic points.  Following the computation
above, we find that this is so for $\Gamma_0(4)$ and $\Gamma^0(4)$.
For $\Gamma_1(3)$ we find that $\phi_* =
\frac{3}{\sqrt{6}}+\frac{i}{2}$ is fixed under $\left(
\begin{array}{cc} 1 & 1 \\ -3 & -2 \end{array}\right)$.

\subsection{Modular anomalies}
\label{app:modularAnomalies}

The modular anomaly coefficients of this model can be computed with the modular weights $n^i,\,\ell^3$
provided in Table~\ref{tab:4DSpectrumModel} by using Eq.~\eqref{eq:modularanomaly}. The result is given
in Table~\ref{tab:modularAnomalies}. Since all elements of the point group act non-trivially in the first
torus, the anomaly coefficients $b'^1_a$ are universal. In contrast, they are different in the second and
third planes for each of the groups.

\begin{table}[!h!]
  \centering
  \begin{tabular}{|l|c|c|c|}
  \hline
   \ $_a$\qquad\ \ $^j$ & 1 & 2 & 3 \\
  \hline\hline
    $\SU3_C$ & $-\tfrac{19}{3}$ & $\tfrac13$      & $3$ \\
    $\SU2_L$ & $-\tfrac{19}{3}$ & $\tfrac13$      & $3$ \\
    $\SO8$   & $-\tfrac{19}{3}$ &$-\tfrac{23}{3}$ & $1$ \\
    $\SU3$   & $-\tfrac{19}{3}$ &$-\tfrac53$      & $3$ \\
   \hline
  \end{tabular}
  \caption{Non-Abelian modular anomaly coefficients $b'^j_a$ of a concrete \Z6--II model. For $j=3$, both 
    K\"ahler--non-Abelian and the complex structure--non-Abelian mixed anomalies coincide, as 
    expected~\cite{Kaplunovsky:1995jw}.}
  \label{tab:modularAnomalies}
\end{table}

\subsection[${\mathcal N}=2$ theories]{Auxiliary $\bs{{\mathcal N}=2}$ theories}
\label{app:N2theories}

In \Z6--II orbifolds the twist is given by $v =\frac{1}{6}
(0,\,1,\,2,\,-3)$. Taking the twist $v_{{}_{\Z2}} = 3 v$ and 
$v_{{}_{\Z3}}=2 v$ leads to two independent ${\mathcal N}=2$
theories. 
These are important for the computation of the stringy
threshold corrections.  Let us study each case separately. 

\paragraph{\bs{\Z2} theory.} In the former case, $v_{{}_{\Z2}}$ gives rise to a \Z2 orbifold with
${\mathcal N}=2$ in which the second torus is left invariant. Embedding the compactification into the
gauge degrees of freedom amounts to taking the shift vector $V_{\Z2} = 3 V^{\SO{10},1}$ and the same
Wilson lines ${\cal A}_5,\,{\cal A}_3$ as in the ${\mathcal N}=1$ orbifold (see Eq.~\eqref{eq:ModelVandW}). Notice that
the Wilson line ${\cal A}_3$ associated with the invariant torus acts non-trivially on physical states.

The resulting ${\mathcal N}=2$ theory has the gauge group
\begin{equation}
  \label{eq:N2Z2gg}
  \SU5\x\SO{10}\x\SU3\x\U1^5\,.
\end{equation}
Omitting the \U1 charges, the quantum numbers of the matter spectrum are given by~\footnote{We list half-hypermutiplets.}
\begin{equation}
  \label{eq:N2Z2spectrum}
 18 (\bs{5},\,\bs{1},\,\bs{1})\oplus 8 (\bs{1},\,\bs{1},\,\bs{3}) \oplus 4 (\bs{1},\,\bs{1},\,\bs{1})
 \oplus c.c.
\end{equation}
It is straightforward now to compute the beta function coefficients, which will be used below. The
general formula is 
\begin{equation}
   (b^j_a)^{{\mathcal N}=2} ~=~ -2 C(G_a) + \sum_{\alpha} T(R_\alpha^a)\,,
\end{equation}
where $R_\alpha^a$ denote the representations w.r.t. the
non-abelian group $G_a$, and
the summation runs over all half-hypermutiplets of the theory. In the present case, for the
non-Abelian groups we obtain
\begin{equation}
  \label{eq:N2Z2Betas}
  (b^2_{\SU5})^{{\mathcal N}=2} = 8\,,\quad 
  (b^2_{\SO{10}})^{{\mathcal N}=2} = -16\,,\quad 
  (b^2_{\SU3})^{{\mathcal N}=2} = 2\,.
\end{equation}

\paragraph{\bs{\Z3} theory.}  $v_{{}_{\Z3}}$ yields a \Z3 orbifold which preserves ${\mathcal N}=2$ and
leaves the third torus untouched.  The embedding into the gauge degrees of freedom requires $V_{\Z3} = 2
V^{\SO{10},1}$ and the same Wilson lines as before.

The resulting ${\mathcal N}=2$ theory has the gauge group
\begin{equation}
  \label{eq:N2Z3gg}
  \SU3_a\x\SU3_b\x\SO8\x\SU3_c\x\U1^6\,.
\end{equation}
Omitting the \U1 charges, the matter spectrum is given by
\begin{eqnarray}
  \label{eq:N2Z3spectrum}
&&6\left[(\bs{3},\,\bs{1},\,\bs{1},\,\bs{1})\oplus (\bs1,\,\bsb{3},\,\bs{1},\,\bs{1})
    \oplus (\bs1,\,\bs{1},\,\bs{1},\,\bs{3})\oplus (\bs1,\,\bs{1},\,\bs{1},\,\bsb{3})\right]
\oplus (\bsb{3},\,\bs{3},\,\bs{1},\,\bs{1})\\
&&\oplus \
3\left[(\bsb{3},\,\bs{1},\,\bs{1},\,\bs{1})\oplus (\bs1,\,\bs{3},\,\bs{1},\,\bs{1})\right]
\oplus 7 (\bs{1},\,\bs{1},\,\bs{8},\,\bs{1})
\oplus 17 (\bs{1},\,\bs{1},\,\bs{1},\,\bs{1}) \oplus c.c.\nonumber
\end{eqnarray}
Hence, the corresponding non-Abelian ${\mathcal N}=2$ beta function coefficients are given by
\begin{equation}
  \label{eq:N2Z3Betas}
  (b^3_{\SU3_a})^{{\mathcal N}=2} = (b^3_{\SU3_b})^{{\mathcal N}=2} = (b^3_{\SU3_c})^{{\mathcal N}=2} = 6 \,,
  \quad (b^3_{\SO{8}})^{{\mathcal N}=2} = 2\,.
\end{equation}

\subsection[Universal $\delta_{GS}^j$]{Universal \bs{\delta_{GS}^j}}
\label{app:deltaGS}

At this point, we have all ingredients to compute $\delta_{GS}^j$ as prescribed by
Eq.~\eqref{eq:relationbeta-delta}. Since our compactification has universal Ka\v{c}-Moody level $k_a=1$,
we must simply use 
\be 
\delta_{GS}^j = b_a^{'j}-\frac{|P_j|}{|P|} (b_a^j)^{{\mathcal N}=2}\,.  
\ee 
For the second torus we have $|P_2|/|P|=1/3$ whereas for the third torus $|P_3|/|P|=1/2$. Furthermore,
one has to be careful with how the 4D gauge group is embedded in the larger $\mathcal{N}=2$ theories. We
find {\it e.g.}~that $\SU2_L$ is a subgroup of \SU5 in the \Z2 subsector, implying that for $\SU2_L$
$\delta_{GS}^2=b'^2_{\SU2_L}-\frac13(b^2_{\SU5})^{{\mathcal N}=2}$. 
One can readily verify then that
\begin{equation}
  \label{eq:deltaGSs}
  \delta_{GS}^1 = -\tfrac{19}{3}\,,\quad \delta_{GS}^2 =-\tfrac73\,,\quad \delta_{GS}^3 =0\,.
\end{equation}
Note that the same $\delta_{GS}^3$ enters the threshold correction contributions containing 
$T_3$ and $U_3$. Besides, $\delta_{GS}^3=0$ holds always for \Z6-II orbifolds
\footnote{This is in general true in torii where the condition $|P|/|P_j|=2$ is satisfied~\cite{Kaplunovsky:1995jw}.}.

\section{Details of Model II}
\label{app:Model2}

In this appendix we describe in detail our second explicit example of an MSSM orbifold compactification, in an analogous way to that above.

\subsection{Model definition}
\label{app:Model2def}

The model is defined by the following shift vector and Wilson lines
\begin{eqnarray}
  \label{eq:Model2VandW}
  V^{\text{SO(10),1}} & = &
    \left(\tfrac13,\,-\tfrac12,\,-\tfrac12,\,0,\,0,\,0,\,0,\,0\,\right) \left( \tfrac12,\,-\tfrac16,\,-\tfrac12,\,-\tfrac12,\,-\tfrac12,\,-\tfrac12,\,-\tfrac12,\,\tfrac12\,\right)\,,\\
 {\cal A}_5 &= &
    \left(-\tfrac{3}{4},\,-\tfrac{1}{4},\,-\tfrac{1}{4},\,-\tfrac{1}{4},\,-\tfrac{1}{4},\,\tfrac{1}{4},\,\tfrac{13}{4},\,\tfrac{13}{4}\,\right) \left( \tfrac{5}{2},\,-\tfrac{3}{2},\,-\tfrac{3}{2},\,-\tfrac{3}{2},\,-\tfrac{3}{2},\,-\tfrac{3}{2},\,-\tfrac{3}{2},\,\tfrac{3}{2}\,\right)\,,\nonumber\\
 {\cal A}_3 &= &
   \left(-\tfrac{1}{2},\,-\tfrac{1}{2},\,\tfrac{1}{6},\,\tfrac{1}{6},\,\tfrac{1}{6},\,\tfrac{1}{6},\,\tfrac{31}{6},\,\tfrac{31}{6}\,\right) \left( 0,\,0,\,0,\,0,\,0,\,0,\,0,\,0\,\right)\,.\nonumber
\end{eqnarray}
The Wilson lines ${\cal A}_3$ (order 3) and ${\cal A}_5$ (order 2) lift the
degeneracy of the fixed points in the second and third torus,
respectively. A possible second order 2 Wilson line in the third torus
is set to zero.  The 4D gauge group after compactification is
\begin{equation}
  \label{eq:4dggModel2}
 \SU{3}_C\x\SU{2}_L\x\U1_Y\x\left[\SO{14}\x\U1^5\right]\,.
\end{equation}
The \U1 generators are chosen to be
\begin{subequations}
\begin{eqnarray}
\label{eq:U1generatorsModel2}
 t_1 & = & t_{\rm X}=\left(0,\,0,\,\tfrac{2}{3},\,\tfrac{2}{3},\,\tfrac{2}{3},\,\tfrac{2}{3},\,\tfrac{2}{3},\,\tfrac{2}{3}\,\right) \left( 0,\,-2,\,0,\,0,\,0,\,0,\,0,\,0\,\right)\,,\\
 t_2 & = & t_{\rm Y}=\left(0,\,0,\,0,\,-\tfrac{1}{2},\,-\tfrac{1}{2},\,\tfrac{1}{3},\,\tfrac{1}{3},\,\tfrac{1}{3}\,\right) \left( 0,\,0,\,0,\,0,\,0,\,0,\,0,\,0\,\right)\,,\\
 t_3 & = &\left(3,\,0,\,0,\,0,\,0,\,0,\,0,\,0\,\right) \left( 0,\,0,\,0,\,0,\,0,\,0,\,0,\,0\,\right)\,,\\
 t_4 & = &\left(0,\,3,\,0,\,0,\,0,\,0,\,0,\,0\,\right) \left( 0,\,0,\,0,\,0,\,0,\,0,\,0,\,0\,\right)\,,\\
 t_5 & = &\left(0,\,0,\,42,\,-3,\,-3,\,-3,\,-3,\,-3\,\right) \left( 0,\,9,\,0,\,0,\,0,\,0,\,0,\,0\,\right)\,,\\
 t_6 & = &\left(0,\,0,\,0,\,9,\,9,\,9,\,9,\,9\,\right) \left( 0,\,15,\,0,\,0,\,0,\,0,\,0,\,0\,\right)\,.  
\end{eqnarray}
\end{subequations}
Again, $t_Y$ and $t_X$ denote respectively the hypercharge and the anomalous \U1 generator.

\subsection{4D spectrum}
\label{app:4DspectrumModel2}

The spectrum in the
untwisted sector includes: the graviton, the axio-dilaton, $S$, the gauge bosons of the gauge group~\eqref{eq:4dggModel2}, 
the K\"ahler moduli $T_i$, $i=1,2,3$, the complex structure modulus $U_3$ (cf.
Section~\ref{sec:ConcreteExample}), and their corresponding supersymmetric partners. In addition, the
spectrum contains the matter fields with non-zero quantum numbers given
in Table~\ref{tab:4DSpectrumModel2}.
They amount to three generations of quarks and leptons plus vector-like exotics with respect to
$\mathcal{G}_\mathrm{SM}$ and hidden matter.

{\footnotesize
\setlength{\LTcapwidth}{0.95\textwidth}
\begin{longtable}{|@{}c@{}|@{\;}l@{}|@{}rrrrrr|@{}rrr|@{}r|l|}
\caption{The spectrum of the \Z6--II Model II.  We denote the fixed points, $f$,  in the $\vartheta^k$
  (un)twisted sector, with their (real) lattice coordinates.  A pair of coordinates $(-,-)$ indicates that the corresponding torus is left fixed
  in the sector.  We give the representations under $SU(3)_C \times SU(2)_L \times SO(14)$ respectively, and the  \U1 charges $q_i$ with respect to the
  basis of generators given above.   $n^i$ and $\ell^i$ are the K\"ahler and complex structure modular weights in the $i$th torus.} 
\label{tab:4DSpectrumModel2}
\\
\hline
\ \ \ $f$ & \ Repr. & $q_\mathrm{X}$ & $q_{Y}$ & $q_{3}$ & $q_{4}$ & $q_{5}$ & $q_{6}$ & $n^1$ & $n^2$ & $n^3$ & $\ell^3$ & \!\!Label\!\! \\
\hline\hline
\endfirsthead
\hline
\ \ \ $f$ & \ Repr. & $q_\mathrm{X}$ & $q_{Y}$ & $q_{3}$ & $q_{4}$ & $q_{5}$ & $q_{6}$ & $n^1$ & $n^2$ & $n^3$ & $\ell^3$ & \!\!Label\!\! \\
\hline\hline
\endhead
\hline
\multicolumn{13}{|r|}{{\it continued ...}$\phantom{_{6}^6}$}\\
\hline
\endfoot
\hline
\endlastfoot
\multicolumn{13}{|l|}{$\vartheta^0$ sector}\\
\hline
 & $(\boldsymbol{1}, \boldsymbol{1}, \boldsymbol{1})$ & $0$ & $1$ & $-\frac{3}{2 }$ & $\frac{3}{2 }$ & $-\frac{45}{2 }$ & $\frac{9}{2 }$ &  $-1$ & $0$ & $0$ &  $0$ & $\bar{e}_{1}$\\
 & $(\boldsymbol{\overline{3}}, \boldsymbol{1}, \boldsymbol{1})$ & $0$ & $-\frac{2}{3 }$ & $-\frac{3}{2 }$ & $\frac{3}{2 }$ & $-\frac{45}{2 }$ & $\frac{9}{2 }$ &  $-1$ & $0$ & $0$ &  $0$ & $\bar{u}_{1}$\\
 & $(\boldsymbol{1}, \boldsymbol{1}, \boldsymbol{1})$ & $0$ & $0$ & $3$ & $3$ & $0$ & $0$ &  $-1$ & $0$ & $0$ &  $0$ & $n_{1}$\\
 & $(\boldsymbol{1}, \boldsymbol{1}, \boldsymbol{14})$ & $2$ & $0$ & $0$ & $0$ & $-9$ & $-15$ &  $0$ & $-1$ & $0$ &  $0$ & $\widetilde{N}_{2}$\\
 & $(\boldsymbol{3}, \boldsymbol{2}, \boldsymbol{1})$ & $0$ & $\frac{1}{6 }$ & $\frac{3}{2 }$ & $-\frac{3}{2 }$ & $-\frac{45}{2 }$ & $\frac{9}{2 }$ &  $0$ & $-1$ & $0$ &  $0$ & $q_{1}$\\
 & $(\boldsymbol{1}, \boldsymbol{2}, \boldsymbol{1})$ & $0$ & $-\frac{1}{2 }$ & $0$ & $0$ & $-45$ & $9$ &  $0$ & $0$ & $-1$ &  $-1$ & $l_{1}$\\
 & $(\boldsymbol{1}, \boldsymbol{2}, \boldsymbol{1})$ & $0$ & $\frac{1}{2 }$ & $0$ & $0$ & $45$ & $-9$ &  $0$ & $0$ & $-1$ &  $-1$ & $\bar{l}_{1}$\\
\end{longtable}
}

\subsection{Trilinear couplings}
\label{app:3pointCouplingsModel2}

The string selection rules allow us to identify all non-vanishing couplings present
in the resulting effective field theory.  In
Table~\ref{tab:3pointCouplingsModel2} we show all couplings allowed by
the standard selection rules; gauge invariance, ${\cal R}$-charge conservation and the space group selection rule.

\begin{table}[!h!]
\begin{center}
{\footnotesize
\begin{tabular}{|c|cc|ccc|c|c|}
\hline
$\vartheta^0\,\vartheta^0\,\vartheta^0$ & \multicolumn{2}{c|}{$\vartheta^3\,\vartheta^4\,\vartheta^5$} 
        & \multicolumn{3}{c|}{$\vartheta^2\,\vartheta^5\,\vartheta^5$} 
        & $\vartheta^2\,\vartheta^4\,\vartheta^0$ & $\vartheta^3\,\vartheta^3\vartheta^0$\\ 
\hline\hline
 $\bar{u}_{1}\ \bar{l}_{1}\ q_{1}$ & $\bar\delta_{1}\ n_{19}\ \delta_{7}$ & $\bar\delta_{3}\ n_{19}\ \delta_{7}$ & $d_{1}\ \bar{e}_{2}\ \bar{u}_{2}$ & $d_{2}\ \bar{e}_{2}\ \bar{u}_{2}$ & $l_{2}\ \bar{e}_{2}\ l_{7}$ & $\bar{l}_{1}\ l_{2}\ n_{19}$ & $q_{1}\ \bar\delta_{1}\ m_{1}$  \\
 & $\bar\delta_{1}\ n_{24}\ \delta_{7}$ & $\bar\delta_{3}\ n_{24}\ \delta_{7}$ & $l_{3}\ \bar{e}_{2}\ l_{7}$ & $l_{2}\ \bar{d}_{6}\ q_{2}$ & $l_{3}\ \bar{d}_{6}\ q_{2}$ & $\bar{l}_{1}\ n_{7}\ l_{5}$ & $q_{1}\ \bar\delta_{2}\ m_{2}$  \\
 & $m_{1}\ n_{18}\ m_{11}$ & $m_{5}\ n_{18}\ m_{11}$ & $d_{1}\ q_{2}\ q_{2}$ & $d_{2}\ q_{2}\ q_{2}$ & $d_{1}\ \bar{d}_{6}\ n_{36}$ & $l_{1}\ n_{5}\ \bar{l}_{3}$ & $q_{1}\ \bar\delta_{3}\ m_{5}$  \\
 & $m_{1}\ n_{23}\ m_{11}$ & $m_{5}\ n_{23}\ m_{11}$ & $d_{2}\ \bar{d}_{6}\ n_{36}$ & $d_{1}\ \bar{e}_{3}\ \bar{u}_{3}$ & $d_{2}\ \bar{e}_{3}\ \bar{u}_{3}$ & $\bar{l}_{1}\ l_{3}\ n_{24}$ & $q_{1}\ m_{3}\ \bar\delta_{4}$  \\
 & $\bar\delta_{2}\ n_{19}\ \delta_{8}$ & $\bar\delta_{5}\ n_{19}\ \delta_{8}$ & $l_{2}\ \bar{e}_{3}\ l_{8}$ & $l_{3}\ \bar{e}_{3}\ l_{8}$ & $l_{2}\ \bar{d}_{7}\ q_{3}$ & $l_{1}\ n_{8}\ \bar{l}_{4}$ & $q_{1}\ \bar\delta_{5}\ m_{9}$  \\
 & $\bar\delta_{2}\ n_{24}\ \delta_{8}$ & $\bar\delta_{5}\ n_{24}\ \delta_{8}$ & $l_{3}\ \bar{d}_{7}\ q_{3}$ & $d_{1}\ q_{3}\ q_{3}$ & $d_{2}\ q_{3}\ q_{3}$ & $\bar{l}_{1}\ n_{17}\ l_{6}$ & $q_{1}\ m_{7}\ \bar\delta_{6}$  \\
 & $m_{2}\ n_{18}\ m_{12}$ & $m_{9}\ n_{18}\ m_{12}$ & $d_{1}\ \bar{d}_{7}\ n_{43}$ & $d_{2}\ \bar{d}_{7}\ n_{43}$ & $n_{4}\ n_{39}\ n_{49}$ & $l_{1}\ \bar{l}_{2}\ n_{30}$ &   \\
\hline
\end{tabular}
\caption{All trilinear level couplings allowed by the standard selection rules in Model II.  We display all
160 couplings, classified according to the twisted sector in which the
involved fields appear.}
\label{tab:3pointCouplingsModel2}
}
\end{center}
\end{table}

\subsection{Modular symmetries}
\label{app:modsymModel2}  Much of the discussion is as in Appendix
\ref{app:modsym}.  
The target space modular symmetry for the first plane is the full
group $SL(2,\mathbb{Z})$, whilst again, Wilson lines break the
symmetry in the second and third planes.  The groups can be computed
to be $\Gamma_1(3)$ for the second plane, $\Gamma_1(2)$ acting on the
K\"ahler modulus of the third plane, and $\Gamma^1(2)$ acting on its
complex structure modulus.  The fixed points of the first and second
plane are as in \ref{app:modsym}.  For $\Gamma_1(2)$, there is a fixed
point at $\phi_* = \frac12 - \frac{i}{2}$, under the transformation  $\left(
\begin{array}{cc} 1 & -1 \\ 2 & -1 \end{array}\right)$.  For
$\Gamma^1(2)$, the transformation $\left(
\begin{array}{cc} 1 & 2 \\ -1 & -1 \end{array}\right)$, leaves the
point $\phi_* = 1 + i$ fixed.

\subsection{Modular anomalies}
\label{app:modularAnomaliesModel2}

The modular anomaly coefficients of this model are computed with the modular weights $n^i,\,\ell^3$
provided in Table~\ref{tab:4DSpectrumModel2} via Eq.~\eqref{eq:modularanomaly}. The result is given
in Table~\ref{tab:modularAnomaliesModel2}.

\begin{table}[!h!]
  \centering
  \begin{tabular}{|l|c|c|c|}
  \hline
   \ $_a$\qquad\ \ $^j$ & 1 & 2 & 3 \\
  \hline\hline
    $\SU3_C$ & $-11$ & $-1$      & $3$ \\
    $\SU2_L$ & $-11$ & $-1$      & $3$ \\
    $\SO{14}$& $-11$ & $-13$     & $-5$ \\
   \hline
  \end{tabular}
  \caption{Non-Abelian modular anomaly coefficients $b'^j_a$ of Model II. For $j=3$, both 
    K\"ahler--non-Abelian and the complex structure--non-Abelian mixed anomalies coincide, as 
    expected~\cite{Kaplunovsky:1995jw}.}
  \label{tab:modularAnomaliesModel2}
\end{table}

\subsection[${\mathcal N}=2$ theories]{Auxiliary $\bs{{\mathcal N}=2}$ theories}
\label{app:N2theoriesModel2}

As in Model I, there are two ${\mathcal N}=2$ theories, which we study separately in the following. 

\paragraph{\bs{\Z2} theory.} The resulting ${\mathcal N}=2$ \Z2 theory has the gauge group
\begin{equation}
  \label{eq:N2Z2ggModel2}
  \SU5\x\SO{16}\x\U1^4\,.
\end{equation}
Omitting the \U1 charges, the quantum numbers of the matter spectrum are given by~\footnote{We list half-hypermutiplets.}
\begin{equation}
  \label{eq:N2Z2spectrumModel2}
 18 (\bs{5},\,\bs{1})\oplus 4 (\bs{1},\,\bs{1}) \oplus c.c.
\end{equation}
The beta function coefficients are given in this case by
\begin{equation}
  \label{eq:N2Z2BetasModel2}
  (b^2_{\SU5})^{{\mathcal N}=2} = 8\,,\qquad 
  (b^2_{\SO{16}})^{{\mathcal N}=2} = -28\,. 
\end{equation}

\paragraph{\bs{\Z3} theory.} The resulting ${\mathcal N}=2$ \Z3 theory has the gauge 
group
\begin{equation}
  \label{eq:N2Z3ggModel2}
  \SU3_a\x\SU3_b\x\SO{14}\x\U1^5\,.
\end{equation}
Omitting the \U1 charges, the matter spectrum is given by
\begin{eqnarray}
  \label{eq:N2Z3spectrumModel2}
&&6\left[(\bs{3},\,\bs{1},\,\bs{1})\oplus (\bs1,\,\bsb{3},\,\bs{1}) \right]
\oplus (\bs{3},\,\bsb{3},\,\bs{1})\\
&&\oplus \
3\left[(\bs{1},\,\bs{3},\,\bs{1})\oplus (\bsb{3},\,\bs{1},\,\bs{1})\right]
\oplus 7 (\bs{1},\,\bs{1},\,\bs{14})
\oplus 17 (\bs{1},\,\bs{1},\,\bs{1}) \oplus c.c.\nonumber
\end{eqnarray}
Hence, the corresponding non-Abelian ${\mathcal N}=2$ beta function coefficients are given by
\begin{equation}
  \label{eq:N2Z3BetasModel2}
  (b^3_{\SU3_a})^{{\mathcal N}=2} = (b^3_{\SU3_b})^{{\mathcal N}=2} = 6 \,,
  \qquad (b^3_{\SO{14}})^{{\mathcal N}=2} = -10\,.
\end{equation}

\subsection[Universal $\delta_{GS}^j$]{Universal \bs{\delta_{GS}^j}}
\label{app:deltaGSModel2}

As for Model I, using the previous ingredients, it is straightforward to compute now $\delta_{GS}^j$. 
The result is 
\begin{equation}
  \label{eq:deltaGSsModel2}
  \delta_{GS}^1 = -11\,,\quad \delta_{GS}^2 =-\tfrac{11}{3}\,,\quad \delta_{GS}^3 =0\,.
\end{equation}
Note that the same $\delta_{GS}^3$ enters the threshold correction contributions containing 
$T_3$ and $U_3$.

\section[\Z6--II Trilinear Coupling Strengths]{\bs{\Z6}--II Trilinear (Twisted) Coupling Strengths}
\label{app:Yukawas}

In this appendix we compute the trilinear coupling strengths for twisted fields, for a generic \Z6-II orbifold.  These results can then be straightforwardly applied to both our explicit models.  Then we describe how to show that the couplings are modular invariant.
The allowed Yukawa couplings between twisted fields are of the kind
$\vartheta^3\vartheta^4\vartheta^5$ and
$\vartheta^2\vartheta^5\vartheta^5$.  Although we use only the latter
in the main text, here we present them both for completeness. 

\subsection{Couplings $\vartheta^3\vartheta^4\vartheta^5$}
\label{app:Yukawas345}

Since $\vartheta^3$ and $\vartheta^4$ leave respectively the second and third torii fixed,
$\mathcal{J}=\{2,3\}$ in Eq.~\eqref{eq:Yuka1}. The coupling strength is then given by
\begin{equation}
\label{eq:CouplingStrength345}
 Y^{34}_{\alpha\beta\gamma}~=~F_{\alpha\beta\gamma}\,g_s \sqrt{t_1}\ N_{345}\ \sum_u 
     \exp\{-2\pi T_1\, u^t  L_1 u \}\,,
\end{equation}
where
\begin{equation}
\label{eq:MandN345}
 L_1 ~=~
\left(
\begin{array}{ccc}
6 && -3\\
-3   && 2
\end{array}
\right)\,,\qquad N_{345}=\frac{2^{17/6}\pi^2}{3^{3/4}\Gamma^3(1/3)}
\end{equation}
and the 2D vector $u$ reduces to the first two coordinates of $u= f_\beta-f_\alpha +\lambda$ with 
$\lambda\in\Lambda$. We have computed all possible coupling strengths, and we find that there are only four different types of couplings. These are defined by the fixed points and thus are characterised by the difference $f_{\alpha\beta} =  f_\beta-f_\alpha $.  
We provide below the explicit forms and leading contributions. 

\bigskip

\noindent
1)  $f_\alpha=(0,0),\,f_\beta=(0,0),\,f_\gamma=(0,0) $. \\
\phantom{.}\quad $f_{\alpha\beta} = (0,0)$.   $F_{\alpha\beta\gamma} = 1$\\
\phantom{.}\quad Dominant contribution: $\exp\{-2\pi u^t L_1 u T_1\}\approx 1 +\ldots$\\
\phantom{.}\quad \Z6--II sample coupling from Model I: $\bar\delta_1\,n_{30}\,\delta_7$
\vskip2mm 

\noindent
2)  $f_\alpha=(\frac12,0),\,f_\beta=(0,0),\,f_\gamma=(0,0)$. \\
\phantom{.}\quad $f_{\alpha\beta} = (-1/2,0)$.   $F_{\alpha\beta\gamma} = \sqrt{3}$\\
\phantom{.}\quad Dominant contribution: $\exp\{-2\pi u^t  L_1 u T_1\}\approx e^{-\pi T_1/2} +\ldots$\\
\phantom{.}\quad \Z6--II sample coupling from Model I: $\bar\delta_3\,n_{30}\,\delta_7$
\vskip2mm 

\noindent
3)  $f_\alpha=(0,0),\,f_\beta=(\frac13,0),\,f_\gamma=(0,0)$. \\
\phantom{.}\quad $f_{\alpha\beta} = (1/3,0)$. $F_{\alpha\beta\gamma} = \sqrt{2}$\\
\phantom{.}\quad Dominant contribution: $\exp\{-2\pi u^t  L_1 u T_1\}\approx e^{-\pi T_1/3} +\ldots$\\
\phantom{.}\quad \Z6--II sample coupling from Model I: $\bar\delta_1\,n_{42}\,\delta_7$
\vskip2mm 

\noindent
4)  $f_\alpha=(\frac12,0),\,f_\beta=(\frac13,0),\,f_\gamma=(0,0)$. \\
\phantom{.}\quad $f_{\alpha\beta} = (-1/6,0)$.  $F_{\alpha\beta\gamma} = \sqrt6$\\
\phantom{.}\quad Dominant contribution: $\exp\{-2\pi u^t  L_1 u T_1\}\approx e^{-\pi T_1/3} +\ldots$\\
\phantom{.}\quad \Z6--II sample coupling from Model I: $\bar\delta_3\,n_{42}\,\delta_7$

\subsection{Couplings $\vartheta^2\vartheta^5\vartheta^5$}
\label{app:Yukawas255}

Since $\vartheta^2$ leaves the third torus invariant,  $\mathcal{J}=\{3\}$ in Eq.~\eqref{eq:Yuka1}.
The coupling strength is then given by
\begin{equation}
 Y^{25}_{\alpha\beta\gamma}~=~F_{\alpha\beta\gamma}\,g_s \sqrt{t_1t_2}\ N_{255}\ \sum_{u_1,u_2} 
     \exp\{-\pi(T_1\, u_1^t L_1 u_1 + T_2\, u_2^t  L_2 u_2)\}
\end{equation}
where $L_1$ is given in Eq.~\eqref{eq:MandN345} and
\begin{equation}
\label{eq:MandN255}
 L_2 ~=~
\left(
\begin{array}{ccc}
2 & &-1\\
-1  & & 2
\end{array}
\right)\,, \qquad N_{255} = \frac{2^{14/3}\pi^4}{3^{3/2}\Gamma^6(1/3)}
\end{equation}
and the 4D vector $u=(u_1,u_2)$ corresponds to the coordinates on the first two torii of
$u= f_\beta-f_\alpha + \lambda$, where $\lambda$ is a lattice vector. 
There are again four types of independent couplings as given below, with their  leading contributions. 
\vskip2mm 

\noindent
1) $f_\alpha=f_\beta=f_\gamma=(0,0,f_{\SU3})$ with  $f_{\SU3}=(0,0)$ or $(\frac13,\frac23)$ or $(\frac23,\frac13)$. \\
\phantom{.}\quad $f_{\alpha\beta} = (0,0,0,0)$. $F_{\alpha\beta\gamma} = 1$\\
\phantom{.}\quad Dominant contribution: $\exp\{-\pi(T_1\, u_1^t  L_1 u_1 + T_2\, u_2^t L_2 u_2)\}\approx 1 +\ldots$\\
\phantom{.}\quad \Z6--II sample coupling from Model I: $\bar{d}_1\,\bar{d}_6\,\bar{u}_2$
\vskip2mm 

\noindent
2)  $f_\alpha=(\frac13,0,f_{\SU3}),\,f_\beta=f_\gamma=(0,0,f_{\SU3})$  with  $f_{\SU3}=(0,0)$ or $(\frac13,\frac23)$ or $(\frac23,\frac13)$. \\
\phantom{.}\quad $f_{\alpha\beta} = (-1/3,0,0,0)$. $F_{\alpha\beta\gamma} = \sqrt{2}$\\
\phantom{.}\quad Dominant contribution: $\exp\{-\pi(T_1\, u_1^t  L_1 u_1 + T_2\, u_2^t L_2 u_2)\}\approx e^{-2\pi T_1/3} +\ldots$\\
\phantom{.}\quad \Z6--II sample coupling from Model I: $\bar{d}_2\,\bar{d}_6\,\bar{u}_2$
\vskip2mm 

\noindent
3)  $f_\alpha=(0,0,\frac13,\frac23),\,f_\beta=(0,0,0,0),\,f_\gamma=(0,0,\frac23,\frac13)$ or permutations thereof.\\
\phantom{.}\quad $f_{\alpha\beta} = (0,0,-1/3,-2/3)$. $F_{\alpha\beta\gamma} = 1$\\
\phantom{.}\quad Dominant contribution: $\exp\{-\pi(T_1\, u_1^t  L_1 u_1 + T_2\, u_2^t L_2 u_2)\}\approx e^{-2\pi T_2/3} +\ldots$\\
\phantom{.}\quad \Z6--II sample coupling from Model I: $n_9\,n_{52}\,n_{64}$
\vskip2mm 

\noindent
4)  $f_\alpha=(\frac13,0,\frac13,\frac23),\,f_\beta=(0,0,0,0),\,f_\gamma=(0,0,\frac23,\frac13)$ or \\
\phantom{.}\quad $f_\alpha=(\frac13,0,\frac23,\frac13),\,f_\beta=(0,0,0,0),\,f_\gamma=(0,0,\frac13,\frac23)$\\
\phantom{.}\quad $f_{\alpha\beta} = (-1/3,0,-1/3,-2/3), (-1/3,0,-2/3,-1/3)$. $F_{\alpha\beta\gamma} = \sqrt{2}$\\
\phantom{.}\quad Dominant contribution: $\exp\{-\pi(T_1\, u_1^t L_1 u_1 + T_2\, u_2^t L_2 u_2)\}\approx e^{-2\pi(T_1+T_2)/3} +\ldots$\\
\phantom{.}\quad \Z6--II sample coupling from Model I: $n_{23}\,n_{52}\,n_{64}$

\subsection{Modular covariance of Yukawas}
\label{app:modinvyuk}
We now illustrate how to check that the Yukawa
couplings transform covariantly under the modular transformations.
Let us first focus on the $\vartheta^2\vartheta^5\vartheta^5$ singlet couplings that we used in the main text for Model I.  These are the couplings $
  n_9 n_{52} n_{64}$,   
$n_{23} n_{52} n_{64}$, $n_{9} n_{59} n_{66}$,  $n_{23} n_{59}
n_{66}$, and they are of kind 3) and 4) above.  We label as  $h_3$ and $h_4$ their respective contributions to
the holomorphic superpotential.   Recalling that fields of the same charges and modular weights can transform amongst themselves under the modular symmetry, it turns out that previous couplings also transform amongst themselves.  Let us now show this explicitly.

Consider the $SL(2,\mathbb{Z})$ 
transformations on $T_1$.  The required transformation property for the couplings is of
the form 
(see Eq. \eqref{eq:YukModularTrafo})
\bea
\left( \begin{array}{c} h_{3} \\ h_{4} \end{array} \right)\rightarrow
\,  (icT_1+d) \, M \, \left(\begin{array}{c} h_{3} \\ h_{4}
\end{array} \right)
\eea
for some field-independent matrix $M$.  Here, we have
used that the
sum over modular weights in the first plane is
$n^1_{\alpha}+n^1_{\beta}+n^1_{\gamma} = -2$ for all the couplings.
It is enough to show that the couplings
transform well under the generators $\mathcal{S},\mathcal{T}$ of
$SL(2,\mathbb{Z})$. 

Take for example the $\mathcal{S}$ transformation on the first torus of a coupling of the third  kind
\bea
&& h_{3}(1/T_1,T_2) = N_{255} \sum_{u_1,u_2} \exp\{-\pi u_1^t \,\frac{L_1}{T_1} \, u_1\} \exp\{-\pi T_2\, u_2^t L_2 u_2)\} 
\eea
  where $u_1 = \lambda$. Concentrating on the $T_1$ dependent exponential only, we now  apply the Poisson
resummation formula 
\bea
&&\sum_{\lambda\in\Lambda} \exp\left\{ -\pi(\lambda+f)^t \hat P (\lambda+f)
  + 2\pi i\epsilon^t(\lambda+f)\right\}\cr
&&\qquad = \frac{1}{V_{\Lambda}\sqrt{\det \hat P}}
\sum_{\lambda^* \in \Lambda^*}\exp\left\{ -\pi (\lambda^*+\epsilon)^t
\hat P^{-1}(\lambda^*+\epsilon)
  - 2\pi i f^t \lambda^*\right\}
\eea
to the sum over $u_1$.  Next split the sum over integers
$(m_1,n_1)\in \Lambda^*$ into a sum over integers $(p_1,n_1)$ with
$m_1=-3p_1-1,-3p_1,-3p_1+1$.  In 
this way, $h_{3}(1/T_1,T_2)$ can be written as a linear
combination of  $h_{3}(T_1,T_2)$ and
$h_{4}(T_1,T_2)$ above, with
coefficients proportional to $i T_1$ as required.

The computation for the congruence subgroup $\Gamma_1(3)$, which acts
on $T_2$ is similar, but note that one needs to apply the Poisson
resummation formula twice.

We can make an analogous analysis for the Model II couplings.  There we considered the couplings $n_7 n_{68} n_{69}$,  
$n_{17} n_{68} n_{69}$,   $n_{7} n_{61} n_{62}$,  $n_{17} n_{61}
  n_{62}$, which are of kind 1) and 2) above.  It turns out that these couplings transform into each other, with the computation following exactly as above.

\end{document}